\documentclass[twocolumn,groupedaddress,superscriptaddress,amsfonts, amssymb, amsmath, caption]{revtex4-1}

\usepackage{mathtools}

\usepackage[noplusnominus,noequal,noasterisk,nospecials,nolessnomore,italic]{mathastext}
\usepackage[stretch=17,shrink=17,step=1,tracking,kerning,final]{microtype}  % enable character protrusion and font expansion

\usepackage{cmap}  % makes ligatures etc. work with PDF viewers (i.e. makes the PDF copyable and searchable)
\everymath{\displaystyle} %% Displays math a full height, even when in-line with text
\DeclareSymbolFont{UPM}{U}{eur}{m}{n}
\DeclareMathSymbol{\partial}{0}{UPM}{"40}
\usepackage{xfrac} %% \sfrac command for side fractions

\usepackage[version=3,arrows=pgf]{mhchem}  %% to typeset chemical formulae easily (using \ce{})

\usepackage[range-phrase=-, range-units=single]{siunitx} %% SI units
\DeclareSIUnit\molar{\mole\per\cubic\deci\metre}
\DeclareSIUnit\Molar{\textsc{m}}
\DeclareSIUnit\kT{$k_{B}T$}
\DeclareSIUnit\bp{}
\usepackage{xr} % No hyper-ref
\usepackage{hyperref}
\usepackage{cleveref} %% Clever reference: http://tex.stackexchange.com/questions/83037/difference-between-ref-varioref-and-cleveref-decision-for-a-thesis
\crefname{equation}{eq.}{eqs.}
\Crefname{equation}{Eq.}{Eqs.}
\definecolor{orange}{rgb}{1,0.5,0}
\usepackage{placeins}

\usepackage[UKenglish]{babel} %% Controls headers and hyphanation patterns
\usepackage[utf8]{inputenc} % Uses Unicode instead of ASCII

\usepackage{paralist}
\setdefaultenum{(i)}{(a)}{(1)}{(A)}
\usepackage{afterpage} % Full page figures and tables
\usepackage{tabularx, booktabs}
\usepackage{longtable}
\newcolumntype{R}{>{\raggedleft\arraybackslash}X}
\newcommand{\ra}[1]{\renewcommand{\arraystretch}{#1}}

\def\ie{{\textit{i.e.~}}}

\newcommand{\etal}{{\textit{et al.~}}}
\newcommand{\etalnospace}{{\textit{et al.}}}
\newcommand{\etalnoperiod}{{\textit{et al}}}

\newcommand{\textcaption}[1]{\textrm{#1}} % Caption text

\def\yieldtoA(#1){-ln(#1/(1-#1))}%

\newcommand{\beginsupplement}{%
    \setcounter{table}{0}
    \renewcommand{\thetable}{S\arabic{table}}%
    \setcounter{figure}{0}
    \renewcommand{\thefigure}{S\arabic{figure}}%
    \setcounter{section}{0}
    \renewcommand{\thesection}{S\arabic{section}}%
    \setcounter{equation}{0}
    \renewcommand{\theequation}{S\arabic{equation}}%
    \setcounter{page}{1}
}

\usepackage{soul}

\begin{document}

\title{Coarse-grained modelling of strong DNA bending II: Cyclization}

\author{Ryan M. Harrison}
\affiliation{Physical \& Theoretical Chemistry Laboratory, Department of Chemistry, University of Oxford, South Parks Road, Oxford OX1 3QZ, United Kingdom}

\author{Flavio Romano}
\affiliation{Physical \& Theoretical Chemistry Laboratory, Department of Chemistry, University of Oxford, South Parks Road, Oxford OX1 3QZ, United Kingdom}

\author{Thomas E. Ouldridge}
\affiliation{Rudolf Peierls Centre for Theoretical Physics, Department of Physics, University of Oxford, 1 Keble Road, Oxford OX1 3NP, United Kingdom}
\affiliation{Department of Mathematics, Imperial College, 180 Queen's Road, London SW7 2AZ, United Kingdom}

\author{Ard A. Louis}
\affiliation{Rudolf Peierls Centre for Theoretical Physics, Department of Physics, University of Oxford, 1 Keble Road, Oxford OX1 3NP, United Kingdom}

\author{Jonathan P. K. Doye}
\affiliation{Physical \& Theoretical Chemistry Laboratory, Department of Chemistry, University of Oxford, South Parks Road, Oxford OX1 3QZ, United Kingdom}

\date{\today}

\begin{abstract} 
DNA cyclization is a powerful technique to gain insight into the nature of DNA
bending. The worm-like chain model provides a good description of small to
moderate bending fluctuations, but some experiments on strongly-bent shorter
molecules suggest enhanced flexibility over and above that expected from the
worm-like chain. Here, we use a coarse-grained model of DNA to investigate the
thermodynamics of DNA cyclization for molecules with less than 210 base pairs.
As the molecules get shorter we find increasing deviations between our computed
equilibrium $j$-factor and the worm-like chain predictions of Shimada and
Yamakawa. These deviations are due to sharp kinking, first at nicks, and only
subsequently in the body of the duplex. At the shortest lengths, substantial
fraying at the ends of duplex domains is the dominant method of relaxation. We
also estimate the dynamic $j$-factor measured in recent FRET experiments. We
find that the dynamic $j$-factor is systematically larger than its equilibrium
counterpart, with the deviation larger for shorter molecules, because not all
the stress present in the fully cyclized state is present in the transition
state. These observations are important for the interpretation of recent
experiments, as only kinking within the body of the duplex is genuinely
indicative of non-worm-like chain behaviour.
\end{abstract}


\maketitle

\section*{Introduction}

As the mechanical properties of DNA play an important role in its biological
capacities, there has been much activity to accurately characterize these properties,
not only in the elastic regime of small to moderate fluctuations but also
for more strongly stressed systems. For example, DNA is found to
overstretch beyond a salt-dependent critical force \cite{Smith96}. Similarly, in
response to twist DNA forms plectonemes beyond a critical buckling superhelical
density \cite{Strick96}. Here, we are interested in the response of DNA to strong
bending. 

The worm-like chain (WLC) model provides a good description of small to
moderate bending fluctuations in DNA
\cite{shore_energetics_1983,Bustamante94,du_cyclization_2005,mazur_dna_2014}.
However, although there is a consensus that for sufficiently strong bending the
stress will be localized within small regions, often termed ``kinks'', much about
this crossover to non-WLC behaviour remains controversial. For example, a
recent review by Vologodskii \etal \cite{vologodskii_strong_2013} highlighted a
number of open questions, including what is the free energy cost of kink
formation, how does the free-energy of a kink depend on bend angle, what is the
critical curvature that causes the double helix to kink?

One way to address these questions is with molecular simulations of a
coarse-grained DNA model, as such simulations are able to directly 
probe the relevant factors.
In this and an accompanying
paper \cite{Harrison_Cohen} we have begun this task for the oxDNA model
\cite{ouldridge_structural_2011,ouldridge_coarse-grained_2012,sulc_sequence-dependent_2012},
which is able to describe well the thermodynamics of DNA hybridization and
basic mechanical properties such as the persistence length and the torsional
modulus \cite{ouldridge_structural_2011} and which has been productively applied to
study a wide variety of biophysical and nanotechnological
systems \cite{doye_coarse-graining_2013}. In the first paper, we addressed the
thermodynamics of kink formation and one particular experimental system, a
``molecular vice'' \cite{fields_euler_2013}, that probed DNA strong bending.
Here, we consider DNA cyclization, in particular looking at the length of DNA
molecules at which duplex kinking begins to play a role in this process.

\subsection*{DNA Cyclization}

DNA cyclization is a convenient model system used to probe DNA bending. Cyclization experiments were first reported in 1966, albeit on 
48\,500 base pair (bp) $\lambda$-DNA \cite{wang_thermodynamic_1966, wang_probability_1966}. In 1981, Shore \etal developed a method to probe the bending of shorter \SIrange{126}{4361}{bp} fragments  \cite{shore_dna_1981}, later noting periodicity in the cyclization efficiency of \SIrange{237}{254}{bp} fragments \cite{shore_energetics_1983}. 

More recently, there has been a particular interest in probing the cyclization of sub-persistence length DNA, to explore whether this regime is accurately described by the WLC model. For example, in 2004 Cloutier \& Widom (C\&W) \cite{cloutier_spontaneous_2004} challenged the conventional wisdom of WLC flexibility established by Shore \etal \cite{shore_dna_1981, shore_energetics_1983}, claiming much greater than predicted cyclization efficiency in \SIrange{93}{95}{bp} DNA fragments. This controversial finding spurred debate on the characteristic length at which DNA cyclization efficiency deviates from the predictions of the WLC model. Despite much experimental \cite{zhang_high-throughput_2003, cloutier_spontaneous_2004, du_cyclization_2005, vafabakhsh_extreme_2012, Le13, le_probing_2014} and theoretical effort \cite{zhang_statistical_2003, yan_localized_2004, travers_dna_2005, wiggins_exact_2005, forties_flexibility_2009, wilson_generalized_2010, vologodskii_strong_2013,sivak_consequences_2011,xu14,taranova14,chen14b,pollak14,Shin15,Naome15,Salari15}, a consensus has not yet been established.

\begin{figure}
\includegraphics[width=8.5cm]{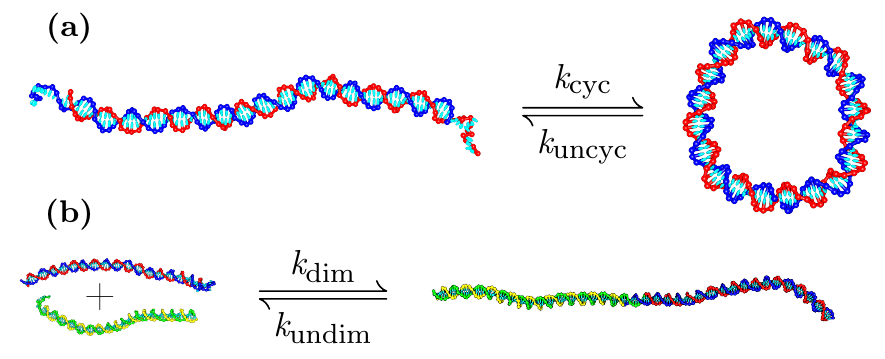}
\caption[]
{
Schematic representations of 
\textbf{(a)} cyclization where $k_\textnormal{cyc}$ and $k_\textnormal{uncyc}$ are the forward and reverse rate constants respectively, and 
\textbf{(b)} dimerization where the rate constants are $k_\textnormal{dim}$ and $k_\textnormal{undim}$. Note that for the dimerization system there is only one complementary sticky end per monomer, the other end being blunt to allow for only one reaction product, a linear dimer.
Figures are oxDNA representations for monomers of length $N_\textnormal{bp}=101$, including complementary sticky ends of length $N_\textnormal{s} = \SI{10}{\bp}$. (Dimer length is $2 N_\textnormal{d} + N_\textnormal{s}$).
}
\label{fig:cyclization-diagram-system-oxDNA}
\end{figure}

A typical cyclization experiment, as depicted in \Cref{fig:cyclization-diagram-system-oxDNA}(a), uses a cyclization substrate with complementary sticky ends, $N_\textnormal{s}$ bases in length, on both ends of a $N_\textnormal{d}$ base-pair duplex. Cyclization leads to the formation of a $N_\textnormal{bp}$-base-pair duplex, where $N_\textnormal{bp} = N_\textnormal{s} + N_\textnormal{d}$. The resultant structure is not a closed minicircle -- two backbone `nicks' are present at either end of the sticky ends. Either the forward rate or equilibrium constant of the cyclization reaction is reported. Experiments differ in how exactly they probe cyclization: methods based on ligation \cite{shore_dna_1981, shore_energetics_1983, cloutier_spontaneous_2004, du_cyclization_2005}, FRET \cite{le_probing_2014, vafabakhsh_extreme_2012,Le13} and multimerization \cite{podtelezhnikov_multimerization-cyclization_2000} have been reported.

In ligase-based experiments, cyclized molecules are fixed in the cyclized state by ligation of the two backbone nicks. The open and ligated cyclized molecules can then be resolved by gel electrophoresis and the concentration of different products measured. FRET-based experiments can be performed in equilibrium, with the molecules allowed to cyclize and uncyclize indefinitely. Fluorophores are attached to both ends of the molecule as FRET reporters: a high FRET signal will be reported when the duplex ends are in close proximity (cyclized), low FRET when apart (open). Although non-WLC behaviour has been suggested by the ligase-based experiments of 
C\&W \cite{cloutier_spontaneous_2004} and the FRET-based experiments of Vafabakhsh \& Ha (V\&H) \cite{vafabakhsh_extreme_2012}, these results have been contested \cite{du_cyclization_2005, vologodskii_strong_2013,vologodskii_bending_2013}. There is not yet a consensus on their interpretation.

Cyclization efficiency is traditionally reported in terms of a $j$-factor, first introduced by Jacobson \& Stockmayer \cite{jacobson_intramolecular_1950}, which is a measure of the effective local concentration of duplex ends at zero end-to-end separation. The $j$-factor enables the ring closure probability to be calculated, and importantly, may be related to a ratio of equilibrium constants:
\begin{equation}
j\textnormal{-factor} = 
j_\textnormal{eq} \equiv
\frac{K_\textnormal{eq}^\textnormal{cyc}}{K_\textnormal{eq}^\textnormal{dim}},
\label{eq:j-eq}
\end{equation}
where $K_\textnormal{eq}^\textnormal{cyc}$ and $K_\textnormal{eq}^\textnormal{dim}$ are the equilibrium constants for cyclization and dimerization, respectively.

Multimerization of the cyclization substrate yields a mixture of linear and
circular products \cite{cloutier_spontaneous_2004,
du_cyclization_2005}. To avoid this complication, a separate dimerization
substrate may be prepared with the same sequence as the cyclization substrate,
but with only one $N_\textnormal{s}$-base complementary sticky end per molecule
\cite{vafabakhsh_extreme_2012}
(\Cref{fig:cyclization-diagram-system-oxDNA}\,(b)). Following hybridization,
the total length of the system is then $2N_\textnormal{d}+N_\textnormal{s}$
base pairs, with blunt as opposed to sticky ends. The consequences of this
choice for $K_\textnormal{eq}^\textnormal{dim}$ and hence $j_\textnormal{eq}$
are discussed in Supplementary
\Cref{sec:cyclization-methods-equilibrium-constants}.

By assuming the contribution from base pairing to $K_\textnormal{eq}^\textnormal{cyc}$ and $K_\textnormal{eq}^\textnormal{dim}$ is the same, the WLC model can be used to estimate $j_\textnormal{eq}$. A common assumption is that the cyclized state is fully stacked, with coaxial stacking across the two nicks. For this situation, the analytic expression derived by Shimada \& Yamakawa (SY) \cite{shimada_ring-closure_1984}, which includes both the bending energy cost of bringing the two ends together and the twist energy cost of bringing the two helix ends into register, is the most appropriate.

Some of the confusion surrounding claims of cyclization efficiency greater than that predicted by the WLC model revolves around the use and interpretation of $j$-factors. While the $j$-factor relation using the ratio of cyclization to dimerization equilibrium constants is well established, experimental studies usually report the ratio of forward rate constants. In the case of ligase-based assays (reviewed in reference \cite{peters_dna_2010}):
\begin{equation}
j_\textnormal{dyn}^\textnormal{ligase} = 
\frac{k_\textnormal{cyc}^\textnormal{ligase}}{k_\textnormal{dim}^\textnormal{ligase}},
\label{eq:j-dyn-ligase}
\end{equation}
where $k_\textnormal{cyc}^\textnormal{ligase}$ and $k_\textnormal{dim}^\textnormal{ligase}$ are the forward rate constants for the formation of the \emph{ligated} circle and dimer, respectively. 

In the experimental limit where the ligation rate is very slow compared to the rate constants for uncyclization ($k_\textnormal{uncyc}$) and undimerization ($k_\textnormal{undim}$), the concentrations of unligated circles and dimers will reach an equilibrium with reactants. If this condition is met, $j_\textnormal{dyn}^\textnormal{ligase}$ should be equivalent to $j_\textnormal{eq}$. In practice, this limit is valid for low ligase concentrations ($\ce{[Ligase]} \ll \ce{[DNA]}$) and short sticky ends \cite{du_cyclization_2005, forties_flexibility_2009}. The importance of this condition is illustrated by Du \etal \cite{du_cyclization_2005}, who suggested that the apparent non-WLC behaviour in the C\&W experiments was rather due to an insufficiently low ligase concentration (reviewed in reference \cite{vologodskii_strong_2013}).

An additional complication with ligase experiments relates to the structure of the substrate, specifically the DNA surrounding the nick. It is unclear whether the ligase will act equally on all nicked duplexes, or only the subset that happen to adopt the right configuration at the nick, be that coaxially stacked or kinked, to allow the ligase to bind. As we show in this work, to alleviate stress, cyclized systems are more likely than dimers to break coaxial stacking at a nick. The ligation rate of hybridized complementary sticky ends may therefore vary substantially depending on the system.

The FRET experiments of V\&H \cite{vafabakhsh_extreme_2012} have the advantage of directly monitoring the transition between cyclized and open states. While it is possible to report thermodynamics from FRET experiments, V\&H report dynamics:
\begin{equation}
j_\textnormal{dyn}^\textnormal{FRET} = 
\frac{k_\textnormal{cyc}}{k_\textnormal{dim}},
\label{eq:j-dyn-FRET}
\end{equation}
where $k_\textnormal{cyc}$ and $k_\textnormal{dim}$ are the forward rate constants for the formation of the \emph{unligated} circle and dimer, respectively. 

Dynamic $j$-factors extracted from FRET-based experiments must also be interpreted with care. Making a comparison between $j_\textnormal{dyn}^\textnormal{FRET}$ and the SY prediction for $j_\textnormal{eq}^\textnormal{WLC}$, V\&H make a claim of much greater than WLC flexibility at $N_\textnormal{bp} \lesssim \SI{100}{bp}$. However, $j_\textnormal{dyn}^\textnormal{FRET} \approx j_\textnormal{eq}$ only in the limit where $k_\textnormal{uncyc} \approx k_\textnormal{undim}$, a condition that V\&H, as well as another more recent FRET experiment \cite{le_probing_2014}, have shown not to be met.
Given that $k_\textnormal{uncyc} \ne k_\textnormal{undim}$, one should not expect $j_\textnormal{dyn}^\textnormal{FRET} \approx j_\textnormal{eq}$ at short $N_\textnormal{bp}$. Thus, the observed deviation of $j_\textnormal{dyn}^\textnormal{FRET}$ from $j_\textnormal{eq}^\textnormal{WLC}$ is not necessarily an indication of non-WLC behaviour.

The interpretation of cyclization experiments is not straightforward; in particular, the microscopic states, responsible for the putative non-WLC flexibility, cannot be directly observed. Additionally, non-WLC behaviour has been reported in a number of other systems, including DNA minicircles 
\cite{du_kinking_2008, demurtas_bending_2009}, a ``molecular vice'' \cite{fields_euler_2013}, and a ``strained duplex'' \cite{qu_elastic_2010, qu_complete_2011,kim15}. As we emphasised in the accompanying paper \cite{Harrison_Cohen}, establishing whether these observations are mutually consistent is a challenging task due to the distinct interplay of mechanics, geometry and topology inherent in each experimental system.

Fortunately, simulation can help bridge this gap. Models that reproduce the basic mechanical, structural and thermodynamic properties of DNA can be used to simulate the relevant systems. The results can be used to establish: (i) whether experimental results are truly indicative of non-WLC behaviour; (ii) whether the behaviour can be attributed to certain types of structure, such as `kinks' \cite{lankas_kinking_2006, mitchell_atomistic_2011, spiriti_dna_2012}; (iii) whether different systems present consistent evidence for non-WLC behaviour; and (iv) whether the inferred occurrence of any disruptions to duplex structure is consistent with our current understanding of DNA thermodynamics.

Here we use oxDNA to probe the cyclization equilibrium as a function of duplex length. This approach provides direct access to microscopic states, enabling a structural interpretation of experimental observations. OxDNA is particularly well-suited to this task, as it provides a good description of both single- and double-stranded DNA, including hybridization thermodynamics, persistence length, torsional modulus and basic structure, and has previously been shown to reproduce a number of stress-induced transitions in DNA \cite{matek_dna_2012,romano_coarse-grained_2013,matek_plectoneme_2014,Mosayebi15}.

\section*{Materials \& Methods}
\subsection*{oxDNA model}

OxDNA
\cite{ouldridge_structural_2011,ouldridge_coarse-grained_2012,sulc_sequence-dependent_2012}
is a nucleotide-level coarse-grained model of DNA 
that has been employed successfully for a wide variety of systems
\cite{doye_coarse-graining_2013}, beginning with the thermodynamic and
structural characterization of DNA nanotweezers \cite{ouldridge_dna_2010}.
Briefly, the model consists of rigid nucleotides with three interaction sites
per nucleotide, interacting via Watson-Crick base pairing, base stacking,
excluded volume, and a potential to represent backbone connectivity. The model
is parameterized to reproduce the thermodynamics of duplex melting at high-salt
(\ce{[Na+]=\SI{500}{\milli\Molar}}), where backbone-backbone electrostatic
repulsion is short-ranged due to counter-ion screening.  Two parameterizations
of the model are available. We use the the average-base
parameterization \cite{ouldridge_structural_2011,ouldridge_coarse-grained_2012},
in which the strength of the base-pairing and stacking interactions are
independent of the identity of the bases, to highlight the basic thermodynamics
of DNA cyclization. By contrast, the sequence-dependent parameterization
\cite{sulc_sequence-dependent_2012} has stacking and base-pairing interactions
that depend on the base identity and is used to compare more directly to the
V\&H experiments.

\subsection*{Simulations}
Simulations of the cyclization equilibrium were performed with a virtual-move Monte Carlo (VMMC) algorithm \cite{whitelam_avoiding_2007} at \SI{298}{\kelvin}. As the free-energy barrier between typical open and cyclized states is large, the transition between the two macrostates constitutes a rare-event. Umbrella sampling, a technique that allows for the biased sampling of states with respect to an order parameter \cite{torrie_nonphysical_1977}, was employed to sample the barrier crossing in reasonable computational time. 

We use a two-dimensional order parameter $Q = (Q_\textnormal{ee}, Q_\textnormal{bp})$ to characterize the transition. $Q_\textnormal{ee}$ is a discretized measure of the distance of closest approach between the complementary sticky ends. $Q_\textnormal{bp}$ is the number of base pairs formed between complementary sticky ends, where $0 \le Q_\textnormal{bp} \le N_\textnormal{s}$. 

To further improve computational efficiency, umbrella sampling was windowed to separately sample the open and cyclized states of each molecule. For the window associated with the open state, the system was restricted to $Q_\textnormal{bp} = 0$; for the window associated with the cyclized state, the system was restricted to $Q_\textnormal{ee} = Q^\textnormal{min}_\textnormal{ee}$ (the value corresponding to the shortest distances between sticky ends). Simulations were run until convergence to within $\pm \SI{5}{\%}$ for each window. 

The sampling windows overlap at $Q = (Q^\textnormal{min}_\textnormal{ee},
Q^{0}_\textnormal{bp})$; results were combined by normalizing each window so
that the free energies were equal for this value of the order parameter. As
there is only one well-defined overlap between the values of the order
parameters for both windows, more complex approaches, such as the weighted
histogram analysis method \cite{kumar_weighted_1992, chodera_use_2007}, were
unnecessary.  To further simplify sampling, we forbade the formation of base
pairs that are not intended in the design of the system (non-native base
pairs). 
Further details of the cyclization simulations can be found in Supplementary
\Cref{sec:cyclization-methods-cyclization-simulations}. The simulation of
dimerization equilibrium is roughly analogous to cyclization, and is elaborated
in Supplementary \Cref{sec:cyclization-methods-dimerization-simulations}. To
compute the equilibrium constants, we deemed all states with $Q_\textnormal{bp}
\ge \SI{1}{\bp}$ to contribute to the cyclized and dimerized states (details in
\Cref{sec:cyclization-methods-equilibrium-constants}).

A complete list of sequences is available in \Cref{tab:cyclization-methods-sequences}. 
Error bars represent the standard error of the mean from 5 independent simulations. 

\subsection*{Structural analysis}
For sufficiently strong bending stress, localized structural disruptions to the DNA double-helical structure are expected. We define three such disruptions, namely fraying, bubble formation and kinking, which are elaborated in detail in the accompanying paper \cite{Harrison_Cohen}. 
Briefly, both fraying and bubble formation involve the breaking of base pairs; the difference is in the location along the duplex. Fraying involves disruption of base pairing at the duplex ends, while bubbles occur in otherwise fully base-paired contiguous stretches away from the duplex ends. 

Conceptually, a kink is an area of strong bending localized to a small segment
of DNA, and can occur both at a nick and within the duplex.  When kinking
occurs in a duplex region, it is nearly always accompanied by bubble formation.
Similarly, kinking at a nick can be accompanied by fraying. We do not attempt
to distinguish different types of kinks, as has been done when analysing
atomistic molecular dynamics simulations \cite{mitchell_atomistic_2011,
zeida_breathing_2012}. 
Instead, we simply define a kink as present within a duplex region when there
is a change in orientation of consecutive bases of greater than 90$^\circ$ on
either strand.  For nicked regions, only consecutive bases on the intact
(unnicked) strand are considered.  Although this cutoff is somewhat arbitrary,
since in the current system, kinks arise to localize bending stress, they are
usually very sharply bent and the criterion works well \cite{Harrison_Cohen}.
Sometimes, however, it gives rise to false negatives, particularly in the case
of kinks at nick sites, and when fraying is present. Nonetheless, it is
satisfactory as an indicator of behaviour for our purposes.
More details, along with subtleties related to kinking at a nick, are discussed
in Supplementary \Cref{sec:cyclization-methods-structural-kink-criterion}.

\begin{figure*}
\includegraphics[width=17.5cm]{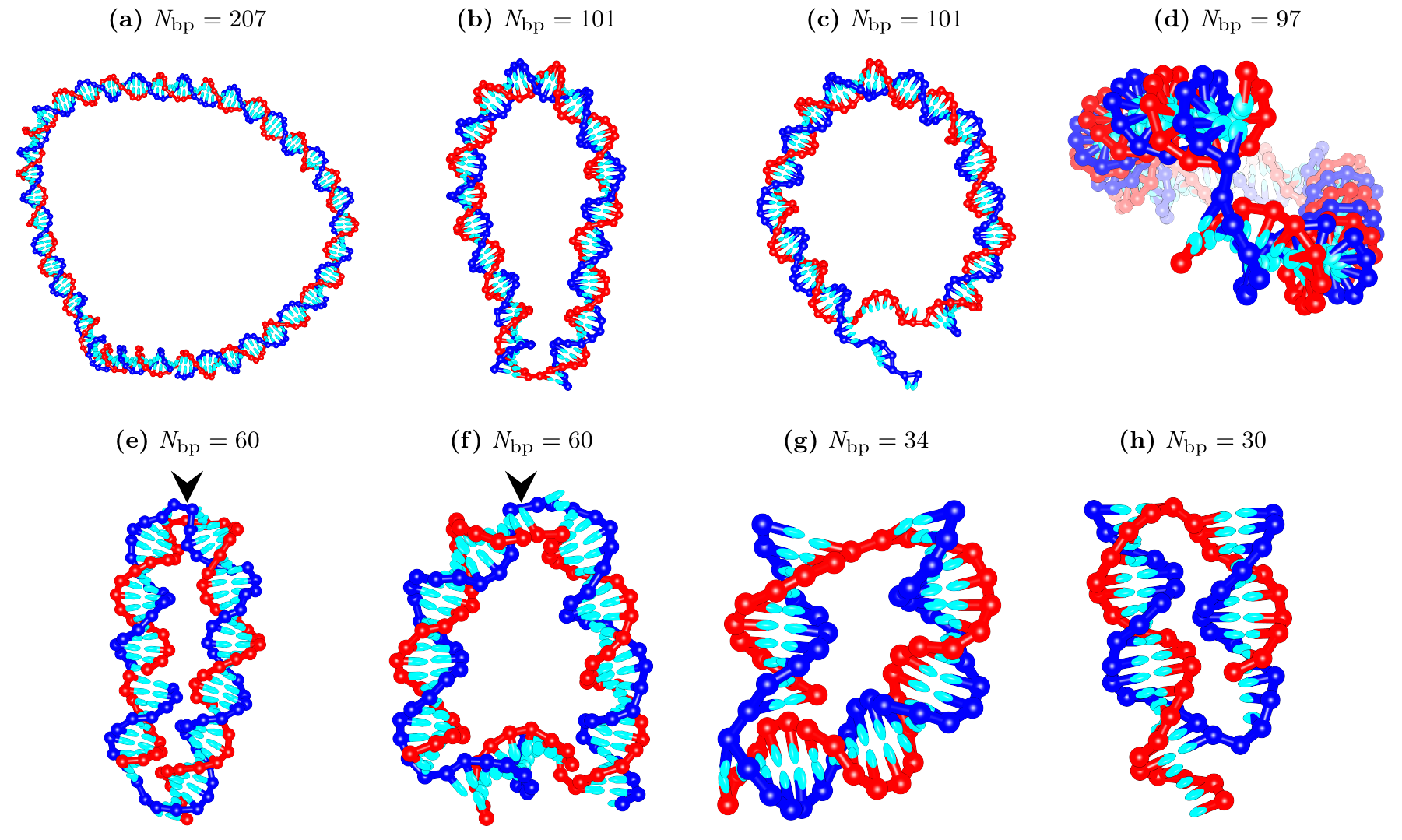}
\caption[]
{
OxDNA representations of different cyclized configurations. Kinks in the duplex, which disrupt stacking and induce a \SIrange{1}{3}{bp} bubble, are indicated with an arrow. All configurations have $N_\textnormal{s}=10$. 
\textbf{(a)} A fully stacked ``circle''.
\textbf{(b)} A ``teardrop'' configuration with a kink at one of the nicks.
\textbf{(c)} A $Q_\textnormal{bp}=1$ ``transition state'' configuration.
\textbf{(d)} Teardrop configurations with $N_\textnormal{bp}\approx(n + \sfrac{1}{2}) \times \textnormal{pitch length}$ can reduce the stress associated with chain continuity at the nick by out-of-plane bending. Configurations with a kink in the duplex and either \textbf{(e)} a kink at one of the nicks or \textbf{(f)} kinks at both nicks. For short duplexes where $N_\textnormal{d}$ is not that much larger than $N_\textnormal{s}$, the sticky ends can associate either by \textbf{(g)} relatively minor bending of the duplex or \textbf{(h)} fraying a few base pairs.
}
\label{fig:cyclization-diagram-zoo}
\end{figure*}

\section*{Results}

We simulate a large range of system sizes; some illustrative configurations are shown in \Cref{fig:cyclization-diagram-zoo}. Quantitatively, we first consider the behaviour of $j_\textnormal{eq}$ as a function of length. For the dimerization system we only computed the equilibrium constant $K_\textnormal{eq}^\textnormal{dim}$ for a few lengths (monomers $N_\textnormal{bp}=30,67,73,101$). As expected, we found $K_\textnormal{eq}^\textnormal{dim}$ to be length-independent to within numerical error (Supplementary \Cref{sec:cyclization-Keq-dimerization}, \Cref{tab:cyclization-Keq-dimerization}); therefore, we use an average value of $K_\textnormal{eq}^\textnormal{dim} = 0.92 \pm \SI{0.20E12}{\per\Molar}$ in our $j$-factor calculation. In contrast, we found $K_\textnormal{eq}^\textnormal{cyc}$, and thereby $j_\textnormal{eq}$ (\Cref{eq:j-eq}), to vary substantially with length.

\begin{figure*}
\includegraphics[width=17.5cm]{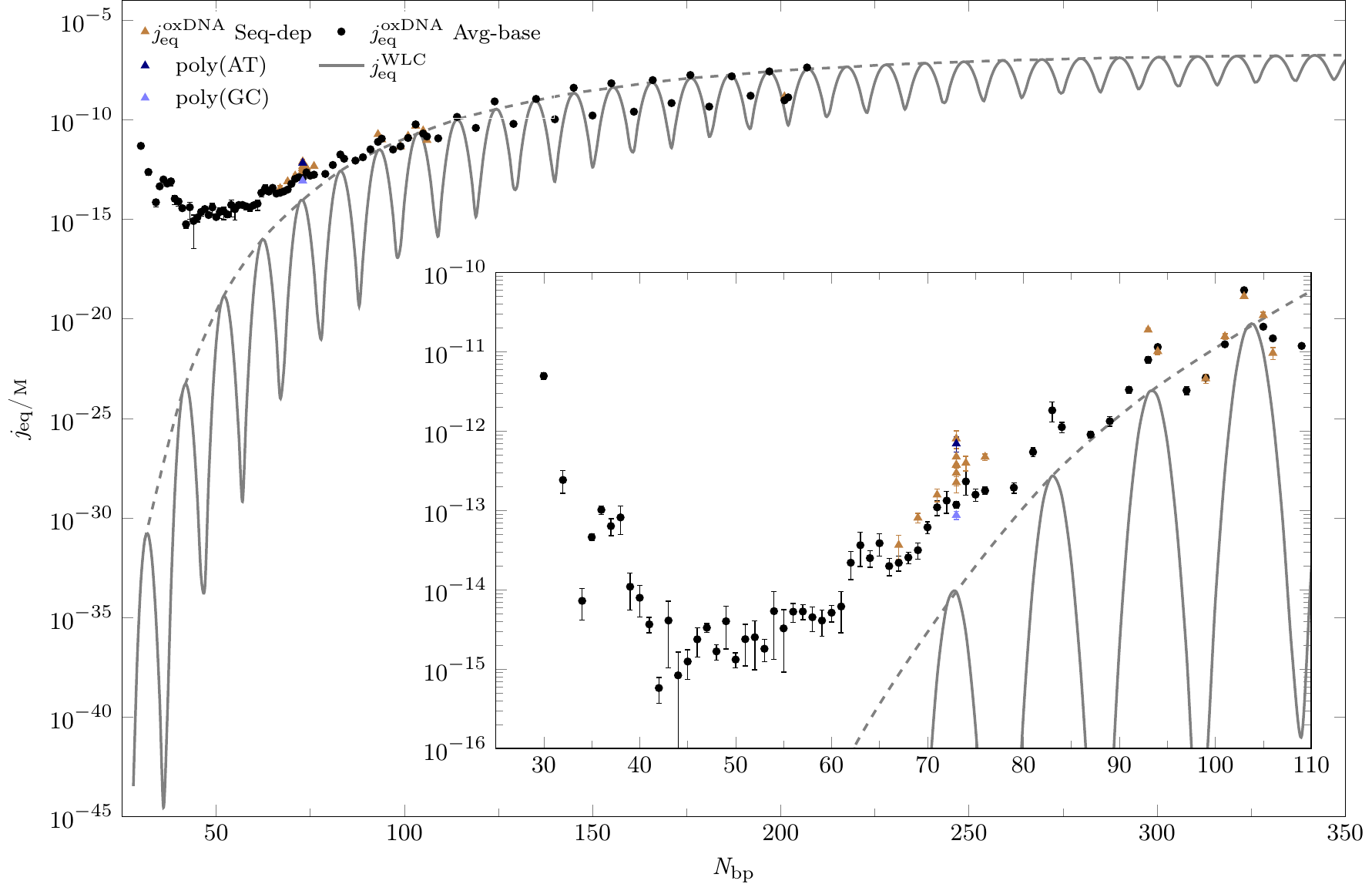}
\caption[]
{
Measured values of 
$j_\textnormal{eq}^\textnormal{oxDNA}$ for the oxDNA average-base parameterization (\textcaption{black circles}) as a function of $N_\textnormal{bp}$ for $N_\textnormal{s} = \SI{10}{\bp}$. For comparison, the Shimada \& Yamakawa (SY) WLC prediction $j_\textnormal{eq}^\textnormal{WLC}$ \cite{shimada_ring-closure_1984} (\textcaption{grey solid line}) is plotted using values of torsional stiffness (\SI{4.75e-28}{\joule\per\meter}), persistence length (\SI{41.82}{\nano\meter}) and pitch length (\SI{10.36}{bp/turn}) appropriate to oxDNA \cite{ouldridge_structural_2011, matek_plectoneme_2014}. The dashed line gives the maxima envelope for the SY prediction (\textcaption{grey dashed line}). Sequence-variation in $j_\textnormal{eq}^\textnormal{oxDNA}$, computed using oxDNA's sequence-dependent parametrization, is illustrated for 
sequences considered by V\&H (\textcaption{brown triangles}), including six at $N_\textnormal{bp}=\SI{73}{bp}$, as well as poly-AT and poly-GC.
}
\label{fig:cyclization-jeq}
\end{figure*}

In \Cref{fig:cyclization-jeq}, we show $j_\textnormal{eq}^\textnormal{oxDNA}$
values, calculated from our measured values of
$K_\textnormal{eq}^\textnormal{cyc}$ and $K_\textnormal{eq}^\textnormal{dim}$
using \Cref{eq:j-eq}, for 81 different lengths in the range $N_\textnormal{bp}
= \SIrange{30}{207}{\bp}$ for fixed $N_\textnormal{s}=\SI{10}{\bp}$ using the
average-base parameterization of oxDNA. These results are compared to
$j_\textnormal{eq}^\textnormal{WLC}$ predictions based on the Shimada \&
Yamakawa (SY) expression \cite{shimada_ring-closure_1984} using previously
calculated values for the relevant structural and mechanical properties of
oxDNA \cite{ouldridge_structural_2011}. 
The SY expression is appropriate for the
formation of a fully stacked ``circle'' configuration with coaxial stacking at
both nicks (\Cref{fig:cyclization-diagram-zoo}\,(a)). Note that the comparison
in \Cref{fig:cyclization-jeq} is fit-free. The effect of varying
$N_\textnormal{s}$ 
as well as the role of nicks and mismatches 
are discussed in Supplementary \Cref{sec:cyclization-experimental-Ha}.

The behaviour of the SY expression is well understood. In the regime of interest, shortening $N_\textnormal{bp}$ tends to make cyclization less favourable as a result of increased bending stress within the duplex. This effect becomes particularly acute for DNA lengths below the persistence length (\SI{41.82}{\nano\meter}/\SI{126}{bp} for the curve plotted in \Cref{fig:cyclization-jeq}). On top of this systematic behaviour, a periodic oscillation is associated with the need to over- or under-twist the duplex when the natural twist is not commensurate with that required to form a closed circle. The magnitude of this oscillation increases at shorter $N_\textnormal{bp}$ because a stronger twist per base pair is required. 

When considering the behaviour of $j_\textnormal{eq}^\textnormal{oxDNA}$ in light of the SY expression, three length-scale dependent regimes become apparent: long ($ N_\textnormal{bp} \gtrsim \SI{80}{\bp}$), intermediate ($N_\textnormal{bp} \approx \SIrange{45}{80}{bp}$) and short ($N_\textnormal{bp} \lesssim \SI{45}{\bp}$). 

In the long length regime ($ N_\textnormal{bp} \gtrsim \SI{80}{\bp}$), oxDNA
reproduces the periodic oscillations predicted by the SY expression, and values
of $j_\textnormal{eq}$ coincide at the maxima of these oscillations. However,
even at the longest 
lengths we consider, the magnitude of
the oscillation is smaller for oxDNA than predicted by the SY expression. At
shorter $N_\textnormal{bp}$, the magnitude of the oscillation decreases for
oxDNA, at odds with the SY expression. The difference stems from the
possibility of adopting alternative ``teardrop'' configurations
(\Cref{fig:cyclization-diagram-zoo}\,(b)), in which most of the twisting
stress, and some of the bending stress, can be relieved by kinking at one of
the nicks, thus breaking coaxial stacking.  In these configurations there is
still the constraint of DNA continuity at the nick and this can now be more
easily satisfied, rather than through over- or undertwist, by out-of-plane
bending (\Cref{fig:cyclization-diagram-zoo}\,(d)). 

The possibility of adopting this alternative teardrop configuration reduces the
free-energy penalty for incommensurate values of $N_\textnormal{bp}$ (i.e.\
$(n + \sfrac{1}{2}) \times \textnormal{pitch length}$), thus
suppressing the oscillations. Further, at shorter $N_\textnormal{bp}$, the
bending stress increases more slowly for these kinked teardrop configurations
than for the coaxially stacked circles. Thus, the difference in
$j_\textnormal{eq}$ between the ``on-register'' (coaxially stacked) and
``off-register'' (kinked at a nick) molecules decreases at shorter
$N_\textnormal{bp}$, rather than increasing as predicted by the SY model.
Although not taken into account for most analyses of cyclization, the
possibility of the cyclized molecule exhibiting a ``teardrop'' configuration
has previously been suggested by Vologodskii \etal
\cite{vologodskii_strong_2013}.

\begin{figure}
\includegraphics[width=8.5cm]{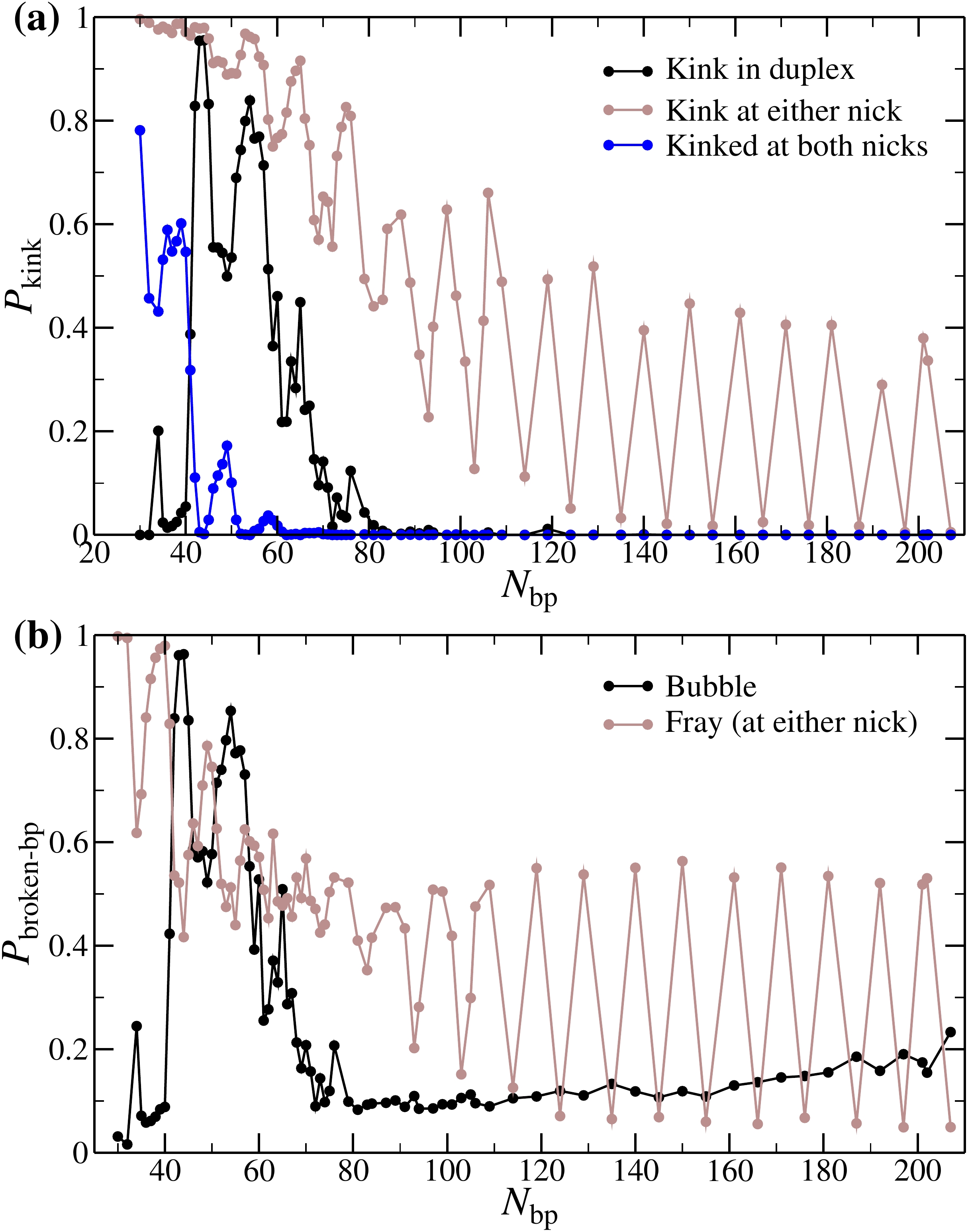}
\caption[]
{
\textbf{(a)} 
Probability of kinking $P_\textnormal{kink}$ as a function of length $N_\textnormal{bp}$, in the duplex, at either nick and at both nicks.
\textbf{(b)} 
Probability of broken base pairs $P_\textnormal{broken-bp}$ as a function of length $N_\textnormal{bp}$ for fraying (base-pairing disruption at either nick) and bubble formation (base-pairing disruption in the duplex region).
}
\label{fig:cyclization-kink}
\end{figure}

Further evidence in support of this analysis is given in \Cref{fig:cyclization-kink}\,(a), which shows the probability of kinking as a function of $N_\textnormal{bp}$. At long lengths ($ N_\textnormal{bp} \gtrsim \SI{80}{\bp}$), kinking does not occur in the duplex regions, but can occur at the nick sites. There is clear periodicity in kinking at a nick.
For example, at $N_\textnormal{bp} = \SI{145}{\bp} \approx 14 \times \textnormal{pitch length}$, the probability of kinking at either of the nicks is 
negligible and the system virtually always adopts a coaxially stacked circle configuration; however, at the longer $N_\textnormal{bp} = \SI{201}{\bp} \approx 19.5 \times \textnormal{pitch length}$, the probability is $\sim \SI{40}{\%}$.
As $N_\textnormal{bp}$ is shortened, the probability of kinking at a nick gradually increases for these ``off-register'' lengths; it is not until $N_\textnormal{bp} = \SI{114}{\bp} \approx 11 \times \textnormal{pitch length}$ that the bending stress along the duplex is sufficient to cause $\sim \SI{10}{\%}$ of molecules to kink at a nick for an ``on-register'' length. 

In the intermediate-length regime ($N_\textnormal{bp} \approx \SIrange{45}{80}{bp}$), we observe enhanced cyclization efficiency compared to the SY prediction. Although $j_\textnormal{eq}^\textnormal{oxDNA}$ continues to decrease with $N_\textnormal{bp}$, it does so more gradually than the SY expression would predict; consequently, $j_\textnormal{eq}^\textnormal{oxDNA}$ is in excess of the peak envelope of the SY expression. 
In this regime, bending stress in the circular coaxially stacked configuration is sufficiently large that kinking at one of the nicks occurs for even the on-register systems (\Cref{fig:cyclization-kink}\,(a)). 

Oscillations due to on-register effects also seem to contribute to $j_\textnormal{eq}^\textnormal{oxDNA}$ in the upper end of the intermediate-length regime, with shallow maxima occurring at $N_\textnormal{bp} = \SI{63}{\bp}$ and $N_\textnormal{bp} = \SI{74}{\bp}$, approximately 6 and 7 times the pitch length. 
However, these sizes no longer correspond to minima in the probability of kinking at a nick (\Cref{fig:cyclization-kink}\,(a)) so the structural underpinnings of these variations is less clear. For shorter lengths, although there are size-dependent variations in $j_\textnormal{eq}^\textnormal{oxDNA}$ (e.g.\ maxima at $N_\textnormal{bp}=43$ and 49), there is no longer a simple relationship to the pitch length, instead reflecting more complex geometric compatibilities that allow cyclized states at these lengths to be particularly stable compared to nearby lengths.

In the intermediate regime, most cyclized molecules are kinked at one of the two nicks. 
At shorter $N_\textnormal{bp}$, the bending stress in the duplex region of the teardrop configurations increases, with the highest curvature being localized opposite the nick that is kinked. Consequently, it becomes increasingly favourable to localize bending stress into a kink in the duplex (unnicked) region, with the probability of this duplex kinking (\Cref{fig:cyclization-diagram-zoo}\,(e)) increasing from near zero at $N_\textnormal{bp} = \SI{81}{\bp}$ to near one at $N_\textnormal{bp} = \SI{43}{\bp}$ (\Cref{fig:cyclization-kink}\,(a)). A kink in the duplex will generally be located opposite a kink at a nick because this arrangement minimizes the residual bending stress in the unkinked portions of the duplex by equalizing the lengths of the double-helical segments between the two kinks. In addition to a loss of stacking, kinking in the duplex typically involves breaking \SIrange{1}{2}{} base pairs \cite{Harrison_Cohen}. A typical configuration for such a kinked duplex state is shown in \Cref{fig:cyclization-diagram-zoo}\,(e). As with kinking at the nick, the availability of this configuration lowers the free-energy cost of cyclization, raising $j_\textnormal{eq}^\textnormal{oxDNA}$ further above the SY prediction.

At the lower end of the intermediate-length regime, the cost of bending without kinking in the duplex region is so high that molecules with duplex kinking dominate the cyclized state. States with two kinks localize the majority of the bending stress at the kinking sites, as shown in \Cref{fig:cyclization-diagram-zoo}\,(e), with long duplex sections relatively relaxed. Consequently, the free-energy cost of looping is largely independent of $N_\textnormal{bp}$ in this regime, causing $j_\textnormal{eq}^\textnormal{oxDNA}$ to level off and reach an approximately constant value. In contrast, the SY $j_\textnormal{eq}^\textnormal{WLC}$ prediction decreases very rapidly. For example, at $N_\textnormal{bp} = \SI{43}{\bp}$, $j_\textnormal{eq}^\textnormal{oxDNA}$ is $10^9$ times greater than the SY prediction.

There exists a rich landscape of structures informed by slight differences in local geometries as a function of $N_\textnormal{bp}$. In addition to the canonical two-kink structures, containing one kink in the duplex and one kink at a nick (\Cref{fig:cyclization-diagram-zoo}\,(e)), we observe several non-trivial arrangements, albeit with relatively low probability. For example, the configuration in \Cref{fig:cyclization-diagram-zoo}\,(f) contains a kink at both nicks as well as in the duplex. 

In the short-length regime ($N_\textnormal{bp} \leq \SI{42}{\bp}$), we observe an increase in $j_\textnormal{eq}^\textnormal{oxDNA}$, in stark contrast to the rapidly decreasing $j_\textnormal{eq}^\textnormal{WLC}$ predicted by the SY expression. Given the many WLC assumptions that are violated at this length scale, a deviation is unsurprising; however, we did not anticipate an increase in $j_\textnormal{eq}^\textnormal{oxDNA}$.

As $N_\textnormal{d}$ is now not that much larger than $N_\textnormal{s}$,
kinking at both nicks allows the single-stranded sticky ends to hybridize
without duplex kinking. The system now generally adopts a conformation of two parallel
duplexes, with the stress now borne by a mixture of continuous bending
(\Cref{fig:cyclization-diagram-zoo}\,(g)) and fraying of a few base pairs at
the ends of the duplexes (\Cref{fig:cyclization-diagram-zoo}\,(h)).
\Cref{fig:cyclization-kink}\,(a) shows a very clear crossover to cyclized
states with kinks at both nicks and no kinks present in the duplex, occurring
abruptly at $N_\textnormal{bp} = \SIrange{40}{42}{\bp}$. At the same time,
fraying at the duplex ends also increases (\Cref{fig:cyclization-kink}\,(b)).
In this regime, as $N_\textnormal{bp}$ is shortened, the difference in length
of the duplexes decreases and the stress in the system tends to drop, leading
to higher $j_\textnormal{eq}$ (\Cref{fig:cyclization-jeq}). Overlaid on this
overall trend are non-trivial geometric effects associated with whether the
lengths of the two duplexes relative to the pitch length is convenient for
connecting them (the thickness of the double helix is now significant compared
to the duplex lengths), which leads to non-monotonic
behaviour of $j_\textnormal{eq}$ and kinking.
Note that 
the value of $N_\textnormal{bp}$ at the crossover between the short- and
intermediate-length regimes is expected to be very dependent on $N_\textnormal{s}$,
occurring at smaller $N_\textnormal{bp}$ for smaller $N_\textnormal{s}$. 

So far we have only reported results using the oxDNA average-base
parameterization, in which the strength of base pairing and stacking
interactions are independent of base identity. As the free-energy cost of
disrupting a duplex to form a kink is sequence-dependent, it is important to
consider how sequence might perturb the general trends we have elaborated thus
far. We therefore studied a variety of sequences used in the V\&H
\cite{vafabakhsh_extreme_2012} experiments including six at $N_\textnormal{bp}
= \SI{73}{\bp}$ using the parameterization of oxDNA that includes
sequence-dependent thermodynamics (\Cref{fig:cyclization-jeq}).

As duplex kinking for oxDNA typically involves the breaking of base pairs, we observe that this kinking preferentially occurs at A--T base pairs. Base pairs that are weaker than average introduce preferred locations for kinking; thus, when duplex kinking is relevant, $j_\textnormal{eq}^\textnormal{oxDNA}$ is expected to be larger for the sequence-dependent than for the average-base parameterization. This is indeed the case for $N_\textnormal{bp} = \SI{73}{\bp}$. The sequence-induced variation for the V\&H $N_\textnormal{bp} = \SI{73}{\bp}$ sequences is a factor of $\sim 4$ (GC-content \SIrange{11}{52}{\%}). This compares to a factor of $\sim 8$ between the extrema in GC-content, poly(AT) and poly(GC).

That we find sequence heterogeneity generally makes duplex kinking easier is
consistent with our explicit investigation of the free energy of duplex kinking
in the accompanying paper \cite{Harrison_Cohen}. 
Our results imply that the crossover to cyclized configurations with duplex
kinking occurs at slightly longer $N_\textnormal{bp}$ when sequence-dependence
is included than for the average-base parameterization.

\begin{figure}
\includegraphics[width=8.5cm]{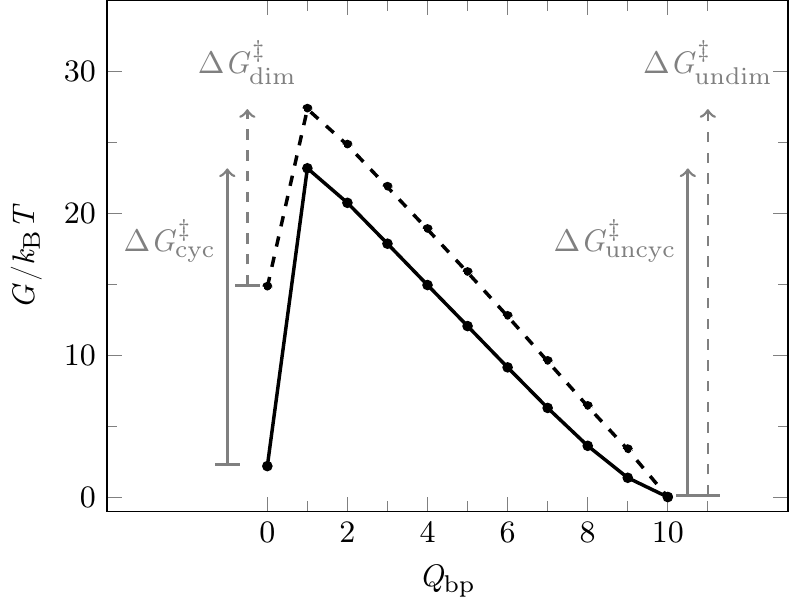}
\caption[]
{
Free energy profiles of cyclization (\textcaption{solid}) and dimerization (\textcaption{dashed}) for a $N_\textnormal{bp}=\SI{101}{\bp}$ system ($N_\textnormal{s}=\SI{10}{\bp}$, $N_\textnormal{d} = \SI{91}{\bp}$). The activation free-energy barriers, $\Delta G^{\ddagger}$ for the forward (cyclization, dimerization) and reverse (uncyclization, undimerization) reactions are labelled. $\Delta G_\textnormal{cyc}^\ddagger$ reflects the free-energy cost of bending to form the first base pair in a cyclization system, whereas $\Delta G_\textnormal{dim}^\ddagger$ reflects the entropic cost of bringing two monomers together within the simulation volume. The dimerization simulations are for a cubic simulation box of dimension \SI{85.18}{\nano\meter}, corresponding to a duplex concentration of $\SI{2.69}{\micro\Molar}$.
Note that for clarity, we have depicted $\Delta G_\textnormal{uncyc}^\ddagger$ and $\Delta G_\textnormal{undim}^\ddagger$ as the free energy difference between the fully base-paired closed state $Q_\textnormal{bp}=10$ and the transition state $Q_\textnormal{bp}=1$; however, in practice, frayed states $Q_\textnormal{bp}=[2,9]$ do contribute to the closed state. While the distinction does not significantly impact our results ($< \sfrac{1}{4}\,\si{\kT}$), we do include the contribution of frayed states in both $\Delta G_\textnormal{uncyc}^\ddagger$ and $\Delta G_\textnormal{undim}^\ddagger$.
}
\label{fig:cyclization-free-energy-cyc-dim}
\end{figure}

As well as equilibrium constants, free-energy profiles as a
function of the number of bases pairs were computed for each system we
considered, as these can give more insight into the pathway for association.
Example profiles
for cyclization and dimerization at $N_\textnormal{bp}=\SI{101}{\bp}$ are
illustrated in \Cref{fig:cyclization-free-energy-cyc-dim}.  One interesting
feature of the profiles is that the free-energy gain from hybridizing the
complementary sticky ends once an initial base pair has formed is less for
cyclization than for dimerization: $\Delta G_\textnormal{uncyc}^\ddagger <
\Delta G_\textnormal{undim}^\ddagger$.  Physically, this indicates that
substantial additional bending stress develops as subsequent base pairs form
for the cyclization system, reducing the free-energy gain upon zippering of the
sticky ends relative to dimerization. For example, the $Q_\textnormal{bp} =
\SI{1}{\bp}$ configuration in \Cref{fig:cyclization-diagram-zoo}\,(c) is
clearly less bent than the $Q_\textnormal{bp} = \SI{10}{\bp}$ configuration in
\Cref{fig:cyclization-diagram-zoo}\,(b)
(Supplementary \Cref{sec:cyclization-diagram-Qbp1}).
The activation energies derivable
from these profiles will be particularly useful in the next section when we
consider the dynamic $j$-factor.

\subsection*{Comparison with experiment}
As noted in the Introduction, the results of ligase experiments performed in
the low ligase concentration limit should, in principle, be comparable to
equilibrium $j$-factors (although there are subtleties related to the ensemble
of states actually detected by the ligation enzymes). Indeed, for long DNA
molecules ($N_\textnormal{bp}$ much longer than the persistence length) and low
ligase concentrations, there has been consistent agreement between experiment
and the WLC model \cite{shore_dna_1981, shore_energetics_1983,
cloutier_spontaneous_2004, du_cyclization_2005, vafabakhsh_extreme_2012,
le_probing_2014}. However, Cloutier \& Widom (C\&W)
\cite{cloutier_spontaneous_2004} reported results for $N_\textnormal{bp}$
shorter than the persistence length ($N_\textnormal{bp} = 93,94,95,105,116$),
showing an apparent deviation from WLC behaviour, with
$j_\textnormal{dyn}^\textnormal{ligase}$ (\Cref{eq:j-dyn-ligase}) enhanced over
the SY WLC prediction $j_\textnormal{eq}^\textnormal{WLC}$
\cite{shimada_ring-closure_1984} by a factor of $10^2 - 10^4$.  The differences
between the maxima in $j_\textnormal{eq}^\textnormal{oxDNA}$ and
$j_\textnormal{eq}^\textnormal{WLC}$ in this size range are much smaller
(\Cref{fig:cyclization-jeq}) and so cannot account for this discrepancy.

In contrast, Du \etal \cite{du_cyclization_2005} found no deviation from WLC
behaviour for $N_\textnormal{bp} = \SIrange{105}{130}{\bp}$. Furthermore, they
presented evidence suggesting that the C\&W experiments used too high a ligase
concentration to enable $j_\textnormal{dyn}^\textnormal{ligase}$ to be compared
with $j_\textnormal{eq}^\textnormal{WLC}$. The results of Du \etal are in good
agreement with the SY WLC expression, albeit with somewhat different materials
parameters (torsional stiffness, persistence length and pitch length) than for
oxDNA, owing in part to different buffer conditions
(Supplementary \Cref{sec:cyclization-experimental-Du-CW},
\Cref{fig:cyclization-experimental-Du-CW}\,(a)).

FRET measurements on DNA cyclization, as pioneered by Vafabakhsh \& Ha (V\&H) \cite{vafabakhsh_extreme_2012}, provide a more direct measure of cyclization because $k_\textnormal{cyc}$, $k_\textnormal{uncyc}$ and $K_\textnormal{eq}^\textnormal{cyc}$ are obtainable, although V\&H mostly report $k_\textnormal{cyc}$. 
V\&H claim enhanced cyclization at $N_\textnormal{bp} \lesssim \SI{100}{\bp}$, based on a comparison between their $j_\textnormal{dyn}^\textnormal{FRET}$ (\Cref{eq:j-dyn-FRET}) and the SY WLC expression $j_\textnormal{eq}^\textnormal{WLC}$. However, as noted earlier, this is only a fair comparison if $k_\textnormal{uncyc} = k_\textnormal{undim}$, which, as $k_\textnormal{undim}$ is expected to be length independent, also implies that $k_\textnormal{uncyc}$ should be independent of $N_\textnormal{bp}$. However, since both V\&H \cite{vafabakhsh_extreme_2012} and more recent FRET measurements \cite{le_probing_2014} suggest that $k_\textnormal{uncyc}$ increases with $N_\textnormal{bp}$, this condition is not met.

Although we do not directly simulate the dynamics of cyclization, we can estimate the relative rates of processes using free-energy profiles such as those in \Cref{fig:cyclization-free-energy-cyc-dim} and unimolecular rate theory for activated processes \cite{hanggi_reaction-rate_1990}. In agreement with experimental investigations, previous work on oxDNA has shown that duplex formation has an effective ``transition state'' involving a very small number of base pairs \cite{ouldridge_dna_2013}. We therefore make the assumption that uncyclization and undimerization rates are given by:
\begin{align}
k_\textnormal{uncyc} &= A \exp{\left( \frac{-\Delta G_\textnormal{uncyc}^\ddagger}{k_\textnormal{B}T} \right)}, \\
k_\textnormal{undim} &= A \exp{\left( \frac{-\Delta G_\textnormal{undim}^\ddagger}{k_\textnormal{B}T} \right)}.
\label{eq:k_uncyc_undim}
\end{align} 
where $A$ is a constant for DNA melting, and $\Delta G_\textnormal{uncyc}^\ddagger$ and $\Delta G_\textnormal{undim}^\ddagger$ are defined in \Cref{fig:cyclization-free-energy-cyc-dim}. The physical content of this assumption is that an increased favourability of base-pair formation is manifested in slower unbinding rates. This is important because bending stress in cyclized systems reduces $\Delta G_\textnormal{uncyc}^\ddagger$ (\Cref{fig:cyclization-free-energy-cyc-dim}). 
The rate constants will be equal ($k_\textnormal{uncyc} = k_\textnormal{undim}$) in the very long length limit ($N_\textnormal{bp}$ much longer than the persistence length). 

\begin{figure}
\includegraphics[width=8.5cm]{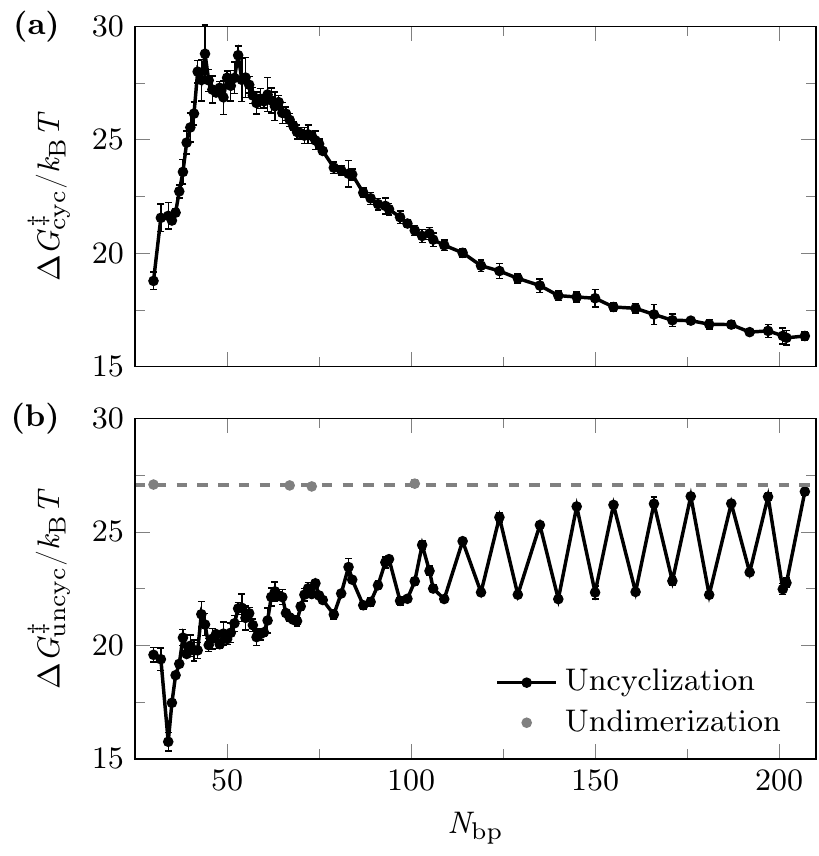}
\caption[]
{
Free-energy barriers for 
\textbf{(a)} cyclization $\Delta G_\textnormal{cyc}^\ddagger$
\textbf{(b)} uncyclization $\Delta G_\textnormal{uncyc}^\ddagger$ (\textcaption{black}) and 
undimerization (\textcaption{grey}, average highlighted with dashed line), 
all computed using the oxDNA average-base parameterization.
}
\label{fig:cyclization-free-energy-stabilization-binding}
\end{figure}

$\Delta G_\textnormal{cyc}^\ddagger$ and $\Delta G_\textnormal{uncyc}^\ddagger$ are plotted as a function of $N_\textnormal{bp}$ in \Cref{fig:cyclization-free-energy-stabilization-binding}. In particular there is a general decrease in $\Delta G_\textnormal{uncyc}^\ddagger$ at shorter lengths, suggesting $k_\textnormal{uncyc}$ increases with shorter $N_\textnormal{bp}$, in agreement with experimental results \cite{vafabakhsh_extreme_2012, le_probing_2014}.  Additionally, any torsional stress in the cyclized state is relieved as base pairs are disrupted, leading to the oscillations in $\Delta G_\textnormal{uncyc}^\ddagger$ at long lengths, with the minima occurring at the more torsionally stressed off-register lengths.

To compare with V\&H's results for $j_\textnormal{dyn}^\textnormal{FRET}$, we note that
\begin{equation}
j_\textnormal{dyn}  = \frac{ k_\textnormal{cyc} }{ k_\textnormal{dim} }
                    = \frac{ K_\textnormal{eq}^\textnormal{cyc} k_\textnormal{uncyc} }{ K_\textnormal{eq}^\textnormal{dim} k_\textnormal{undim} }.
\label{eq:jdyn-from-eq-step1}
\end{equation}

Therefore, using \Cref{eq:k_uncyc_undim}, our approximation for $j_\textnormal{dyn}^\textnormal{oxDNA}$ is
\begin{equation}
j_\textnormal{dyn}^\textnormal{oxDNA} = \frac{ K_\textnormal{eq}^\textnormal{cyc}  \exp \left( \frac{-\Delta G_\textnormal{uncyc}^\ddagger}{k_\textnormal{B}T} \right) }{ K_\textnormal{eq}^\textnormal{dim}  \exp \left( \frac{-\Delta G_\textnormal{undim}^\ddagger}{k_\textnormal{B}T} \right) },
\label{eq:jdyn-from-eq-final}
\end{equation}
at each $N_\textnormal{bp}$.

\begin{figure*}
\includegraphics[width=17.5cm]{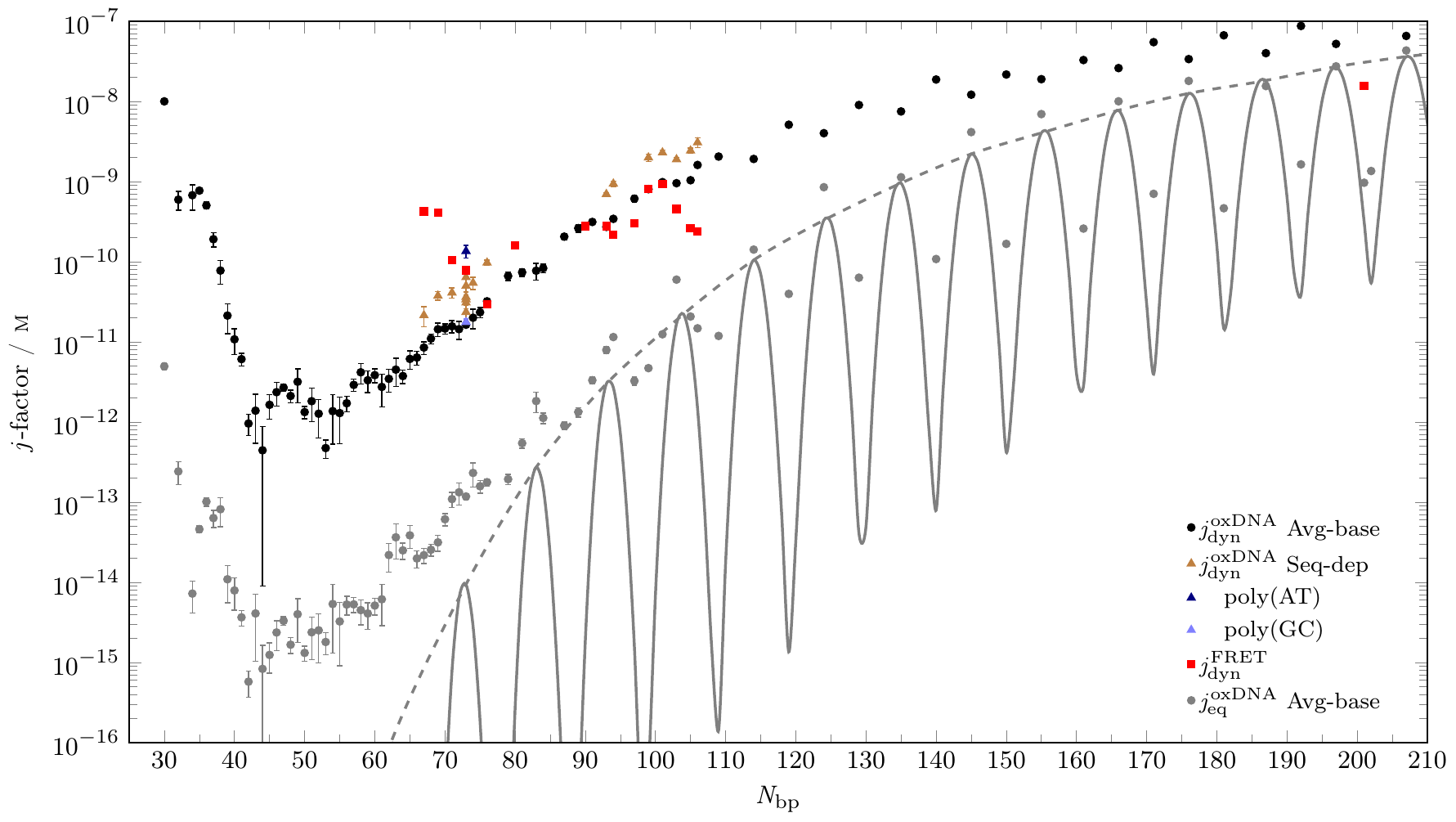}
\caption[]
{
OxDNA dynamic $j$-factor $j_\textnormal{dyn}^\textnormal{oxDNA}$ (\textcaption{black circle}) compared to the FRET experiments of V\&H $j_\textnormal{dyn}^\textnormal{FRET}$ ({\color{black}\textcaption{red squares}}) \cite{vafabakhsh_extreme_2012}. Results for the oxDNA sequence-dependent parameterization using V\&H 14 variable $N_\textnormal{bp}$ sequences, in addition to 6 sequences at $N_\textnormal{bp}=\SI{73}{\bp}$ to highlight the role of sequence-variation. For reference, $j_\textnormal{eq}^\textnormal{oxDNA}$ and $j_\textnormal{eq}^\textnormal{WLC}$ are also shown ({\color{black}\textcaption{grey}}). 
}
\label{fig:cyclization-jdyn}
\end{figure*}

We are now in a position to estimate $j_\textnormal{dyn}^\textnormal{oxDNA}$ from oxDNA's equilibrium constants and activation free-energy barriers. We expect $K_\textnormal{eq}^\textnormal{dim}$ and $\Delta G_\textnormal{undim}^\ddagger$ to be length-independent as the excluded volume of the duplex far from the complementary single-stranded sticky ends is likely to have little effect on the dimerization process. Indeed, this appears to be the case to within less than \SI{0.5}{\kT} in $\Delta G$ (\Cref{fig:cyclization-free-energy-stabilization-binding}(b), Supplementary \Cref{tab:cyclization-Keq-dimerization}). Our resulting $j_\textnormal{dyn}^\textnormal{oxDNA}$ values are plotted in \Cref{fig:cyclization-jdyn}, where they are compared to our $j_\textnormal{eq}^\textnormal{oxDNA}$ values, and V\&H's $j_\textnormal{dyn}^\textnormal{FRET}$.

We observe that oxDNA's $j_\textnormal{dyn}^\textnormal{oxDNA}$ values lie
substantially above $j_\textnormal{eq}^\textnormal{oxDNA}$, except in the limit
of long $N_\textnormal{bp}$ for on-register lengths. Only if all the
stress-induced destabilization of the fully cyclized state is exhibited in the
forward rate would $j_\textnormal{dyn}=j_\textnormal{eq}$. Although the
activation free-energy for cyclization increases as $N_\textnormal{bp}$
decreases (\Cref{fig:cyclization-free-energy-stabilization-binding}(a)), the
full bending and twisting stress present in the cyclized state is not yet
present in the transition state where the two complementary sticky ends have
formed their first base pair, e.g.\ \Cref{fig:cyclization-diagram-zoo}(c), and
so $j_\textnormal{dyn}^\textnormal{oxDNA} >
j_\textnormal{eq}^\textnormal{oxDNA}$.  Instead, stress which is not present in
the transition state leads to a decrease in the free-energy barrier for
uncyclization (\Cref{fig:cyclization-free-energy-stabilization-binding}(b)) and
accelerates uncyclization relative to undimerization. 
Alternatively, one can consider that the stress in the cyclized state subjects the
duplex formed between the sticky ends to a shear force \cite{le_probing_2014}
which is well known to lead to a more rapid rupture of a duplex \cite{Hatch08,Mosayebi15}.

The observed behaviour is
similar to that seen by V\&H; indeed, $j_\textnormal{dyn}^\textnormal{oxDNA}$
and $j_\textnormal{dyn}^\textnormal{FRET}$ agree remarkably well in the range
$N_\textnormal{bp}=\SIrange{90}{105}{\bp}$, with reasonable agreement extending
to $N_\textnormal{bp}=\SI{70}{\bp}$. Our results suggest that variation of
uncyclization rates with $N_\textnormal{bp}$ may be an important contribution
to apparent non-WLC behaviour in $j_\textnormal{dyn}^\textnormal{FRET}$.

To begin to understand the difference between V\&H's dynamic $j$-factor and
$j_\textnormal{eq}^\textnormal{WLC}$ a number of authors have tried to account
for the role of the single-stranded tails in cyclization by incorporating a
``capture radius'' into the WLC $j$-factor calculation, as an approximation for
how close the duplex ends must be in order for the sticky ends to hybridize
\cite{vafabakhsh_extreme_2012, vologodskii_strong_2013,
vologodskii_bending_2013, le_probing_2014}.  Indeed, this approach leads to a
significantly enhanced $j$-factor when a value that is taken to be roughly appropriate
for the 10-base sticky ends of the V\&H experiment, namely \SI{5}{nm}, is used
\cite{vafabakhsh_extreme_2012,vologodskii_bending_2013}.  

Assuming this
approach is an attempt to capture the weaker constraints at the transition
state, and hence to estimate the activation free-energy barrier relevant to
$j_\textnormal{dyn}$ (note $j_\textnormal{dyn}$ and $j_\textnormal{eq}$ are
often not clearly differentiated in discussions of V\&H's results), we can use
our oxDNA results to test the reasonableness of this approximation by measuring
the separation of the duplex ends in the transition state ensemble. At long
lengths our measured capture radius is approximately constant with a value
just under \SI{4}{nm}, but increases at shorter lengths because some of the stress 
at the transition state is partitioned into stretching the single-stranded tails,
reaching a maximum of about \SI{7}{nm} at about $N_\textnormal{bp}\approx 40$
(Supplementary 
\Cref{fig:cyclization-capture-radius}).  
Thus, although the \SI{5}{nm} value used previously \cite{vafabakhsh_extreme_2012,vologodskii_bending_2013} 
is not unreasonable, unsurprisingly this approach does not capture the full
complexity of the transition state to cyclization,
and nor does it account for the strain present in the sticky 
ends for them to achieve contact. 

Therefore, although explaining a dynamic $j$-factor in terms of a capture
radius is physically well-motivated, it does not provide a full understanding.
Moreover, as its value is not tightly constrained, and certain contributions to
stability are neglected, it is difficult to judge whether WLC behaviour is
violated or not by comparing a calculated $j$-factor curve to an experimental
$j_\textnormal{dyn}$.

We note that $j_\textnormal{dyn}^\textnormal{oxDNA}$, in contrast to
$j_\textnormal{eq}^\textnormal{oxDNA}$, varies comparatively smoothly with
$N_\textnormal{bp}$, with only a weak periodicity on the length scale of the
pitch length. The shallow maxima at larger lengths occur at
$N_\textnormal{d}=(n+\sfrac{1}{2})\times \textnormal{pitch length}$, because
the two sticky ends are then on the same side of a torsionally unstressed
duplex.  Consistent with this smooth variation, $\Delta
G_\textnormal{cyc}^\ddagger$ also varies relatively smoothly with
$N_\textnormal{bp}$. The strong periodicity in
$j_\textnormal{eq}^\textnormal{oxDNA}$ comes from the free-energy gain when
zippering up the complementary sticky ends (\ie $\Delta
G_\textnormal{uncyc}^\ddagger$), which is greater when $N_\textnormal{bp}$ is
an integer multiple of the pitch length, allowing the formation of a relatively
relaxed coaxially stacked circle. V\&H suggest that their
$j_\textnormal{dyn}^\textnormal{FRET}$ data at $N_\textnormal{bp} =
\SIrange{93}{106}{\bp}$ displays a strong oscillation with a period of about
one pitch length (\Cref{fig:cyclization-jdyn}). In agreement with Vologodskii
\etalnospace, \cite{vologodskii_strong_2013} we find no physical mechanism for
such a strong oscillation and would suggest that the experimental evidence for
this oscillation is not compelling.

At the shortest lengths investigated, $j_\textnormal{dyn}^\textnormal{FRET}$ still lies above $j_\textnormal{dyn}^\textnormal{oxDNA}$. Some of this difference might be due to our use of the oxDNA average-base parameterization. Sequence does play a role in DNA flexibility, as noted in the accompanying paper \cite{Harrison_Cohen}, but the impact of sequence variation is non-trivial. In particular, because kinks tend to localize to AT base pairs, kinking in the duplex is easier for a sequence with \SI{50}{\%} GC-content than reported for our average-base parametrization. 

Using the oxDNA parameterization with sequence-dependent thermodynamics 
we found $j_\textnormal{dyn}$ to increase compared to the values for the average-base parameterization. Consistent with our results for $j_\textnormal{eq}^\textnormal{oxDNA}$, we find a sequence-induced variation in $j_\textnormal{dyn}^\textnormal{oxDNA}$ of a factor of $\sim 4$ for the six V\&H $N_\textnormal{bp} = \SI{73}{\bp}$ sequences (GC-content \SIrange{11}{52}{\%}) \cite{vafabakhsh_extreme_2012} compared to a factor of $\sim 8$ between the extrema in GC-content, poly(AT) and poly(GC). 
This compares to a factor of $\sim 60$ for experimental looping rates, a discrepancy which may be explained by V\&H's use of poly(A) tracts, a sequence-motif well-known to introduce intrinsic curvature in duplex DNA \cite{rivetti_polymer_1998}. As oxDNA's sequence-dependent parameterization is based on the nearest-neighbour thermodynamics of SantaLucia \etal \cite{santalucia_unified_1998, santalucia_thermodynamics_2004}, alternative structural motifs such as poly(A) tracts are outside the scope of the model. OxDNA also does not reproduce sequence-dependent structural (e.g.\ the difference in size between purine and pyrimidine) or mechanical (e.g.\ flexibility) properties.

The origin of the remaining discrepancy between $j_\textnormal{dyn}^\textnormal{FRET}$ and  $j_\textnormal{dyn}^\textnormal{oxDNA}$ at the shortest $N_\textnormal{bp}$ is not yet clear, but may indicate enhanced flexibility for V\&H versus oxDNA due to WLC (\ie lower persistence length) or non-WLC behaviour (\ie kinking within the duplex at slightly longer $N_\textnormal{bp}$). The oxDNA persistence length of \SI{41.82}{\nano\meter} is within the range of experimental observations at \ce{[Na+]=\SI{500}{\milli\Molar}}, but smaller values at \ce{[Na+]=\SI{750}{\milli\Molar}} are not implausible  \cite{savelyev_monovalent_2012,HerreroGalan13}. It is also possible that oxDNA slightly underestimates the prevalence of kinking within duplex regions \cite{Harrison_minicircles}; an onset of kinking at slightly lower  stress (longer $N_\textnormal{bp}$) would make cyclization at shorter lengths more favourable.  Finally, it is worth noting that the presence of fluorophores may cause perturbations in the V\&H experiments.

V\&H do report $K_\textnormal{eq}^\textnormal{cyc}$ for some systems; oxDNA results are in good agreement for those lengths (Supplementary \Cref{fig:cyclization-experimental-Ha-Ns}).

\section*{Conclusion}
Cyclization is a system dependent manifestation of the general thermodynamics of strong DNA bending, elaborated in the accompanying paper \cite{Harrison_Cohen}. The remarkable range of behaviour in cyclized systems is explicable by the interplay between three specific deformation modes of stressed duplexes: continuous bending, kinking and fraying. 

OxDNA reveals that each of these modes is present at a characteristic length-scale with respect to cyclization: continuous bending at long lengths ($ N_\textnormal{bp} \gtrsim \SI{80}{\bp}$), duplex kinking at intermediate lengths ($N_\textnormal{bp} \approx \SIrange{45}{80}{bp}$) and fraying at short lengths ($N_\textnormal{bp} \lesssim \SI{45}{bp}$). In addition, as $N_\textnormal{bp}$ is shortened, there is an increase in kinking at the two nicks that remain after the hybridization of the sticky ends. At longer lengths, kinking at a nick is only observed for ``off-register'' molecules that cannot form torsionally relaxed coaxially stacked circles. The ability of said nicks to relax bending, as well as torsional stress, means that they become increasingly prevalent for shorter $N_\textnormal{bp}$. At the shortest lengths, kinking at both nicks is dominant.

We use oxDNA to probe the reported observation of non-WLC behaviour in FRET-based cyclization experiments \cite{vafabakhsh_extreme_2012}. In agreement with experiment, we observe that for shorter values of $N_\textnormal{bp}$, the apparent $j$-factor lies substantially above the predictions of the Shimada \& Yamakawa (SY) \cite{shimada_ring-closure_1984} WLC model. We also observe that the periodic oscillations predicted by the SY model are suppressed. This behaviour arises from two conceptually distinct phenomena.

Firstly, highly stressed cyclized systems can adopt configurations that relax stress more effectively than through continuous bending, thereby reducing the overall free-energy cost of cyclization relative to a direct estimate based on a simple WLC-based model. At various values of $N_\textnormal{bp}$, oxDNA identifies kinking at nicks, kinking within the duplex region and fraying of base pairs as key relaxation modes.

Secondly, oxDNA suggests that not all of the reduction in $K_\textnormal{eq}^\textnormal{cyc}$ relative to $K_\textnormal{eq}^\textnormal{dim}$ is due to stress manifest in the cyclization rate; uncyclization rates are also substantially increased relative to undimerization rates. The result is that dynamic $j$-factors based on the ratio of cyclization and dimerization rates lie even further above the SY prediction than their equilibrium $j$-factor counterpart.

Of the above effects, only kinking within the duplex can reasonably be described as truly non-WLC behaviour. WLC and related statistical models do not predict absolute rates directly. Kinking at nicks and fraying can only occur when the DNA backbone is discontinuous, and any resultant effects are unrelated to whether WLC models accurately describes the body of the DNA duplex. 
In oxDNA (with the average-base parameterization), kinking in the duplex only has a substantial effect for $N_\textnormal{bp} \lesssim \SI{70}{bp}$. Note, if sequence dependence is taken into account, then kinking may occur for slightly longer lengths.

Our results suggest that much of the apparent ``extreme bendability'' reported by Vafabakhsh and Ha \cite{vafabakhsh_extreme_2012} can be attributed to factors that are not strictly speaking evidence of non-WLC behaviour. We cannot account for deviations at their very shortest lengths ($N_\textnormal{bp} \lesssim \SI{70}{\bp}$). In oxDNA, kinking within the duplex region is present, but not completely dominant. It is possible that oxDNA slightly overestimates the difficulty of kinking within a duplex -- if this is the case, the data for the very smallest values of $N_\textnormal{bp}$ studied by V\&H may be indicative of duplex flexibility over and above that predicted by the WLC, although we note that to give a substantial effect on $j_\textnormal{eq}$, kinking must not only be present, but must dominate the ensemble. Kinking dominating the ensemble at $N_\textnormal{bp} \approx \SI{70}{\bp}$ is inconsistent with oxDNA predictions, and available experimental evidence for a ``molecular vice'' \cite{Harrison_Cohen} and DNA minicircles \cite{Harrison_minicircles}.

Various authors have incorporated a ``capture radius'' into the WLC $j$-factor calculation to capture phenomenologically some of the effects listed above \cite{vafabakhsh_extreme_2012, vologodskii_strong_2013, le_probing_2014}. Indeed, this approach leads to an enhanced $j$-factor, but the choice of capture radius is somewhat arbitrary and imprecise, making it difficult to assess whether non-WLC behaviour is present. 
Nonetheless, we find that the \SI{5}{\nano\meter} capture radius that has been used in the interpretation of 
V\&H's experiments is not unreasonable. 

OxDNA is only a model, and good correspondence with experimental results should
not be over-interpreted. Nonetheless, the relaxation mechanisms identified are
clearly physically plausible. For example, enhanced uncyclization rates have
previously been noted in the literature \cite{vafabakhsh_extreme_2012,
le_probing_2014} (and also for transcription-factor mediated looping
\cite{Chen14}). It is clear that much of the apparent discrepancy between the
data of V\&H and the predictions of WLC-based models are due to effects that
are not true violations of the WLC model of duplex DNA flexibility. Our study
also helps to reconcile the results of V\&H with previous ligase-based assays
which saw no evidence of enhanced flexibility at $N_\textnormal{bp} \approx
100$ \cite{du_cyclization_2005}, and previous studies of minicircles which
detected no evidence of duplex disruption at these length scales
\cite{du_kinking_2008}.  

To explore whether the shortest lengths studied by V\&H do show evidence of kinking and enhanced flexibility, we would propose experiments of shorter sequences and systematic collection of both dynamic and equilibrium data. The latter is extremely important; statistical WLC models make equilibrium predictions, and so the breakdown of a WLC description can only be confirmed with equilibrium data. Indeed, elucidating the subtleties of dynamic and quasi-dynamic (C\&W) $j$-factors is one of the key issues addressed in this work.

%\section*{Supplementary data}
%Supplementary Data are available at NAR Online.

\section*{Funding}
This work was supported by the Engineering and Physical Sciences Research Council [EP/I001352/1], the National Science Foundation Graduate Research Fellowship Program, the National Institutes of Health National Heart, Lung and Blood Institute, Wolfson College, Oxford, and University College, Oxford.

%\textit{Conflict of interest statement}. None declared.

\section*{Acknowledgements}
The authors acknowledge the computing facilities of the Oxford Advanced Research Computing and the e-Infrastructure South IRIDIS High Performance Computing Facility.

\clearpage

\begin{widetext}
\begin{center}
\textbf{\large 
Supplementary material for ``Coarse-grained modelling of strong DNA bending II: Cyclization''}
\end{center}
\end{widetext}

%\author{Ryan M. Harrison}
%\affiliation{Physical \& Theoretical Chemistry Laboratory, Department of Chemistry, University of Oxford, South Parks Road, Oxford, UK, OX1 3QZ}
%
%\author{Flavio Romano}
%\affiliation{Physical \& Theoretical Chemistry Laboratory, Department of Chemistry, University of Oxford, South Parks Road, Oxford, UK, OX1 3QZ}
%
%\author{Thomas E. Ouldridge}
%\affiliation{Rudolf Peierls Centre for Theoretical Physics, Department of Physics, University of Oxford, 1 Keble Road, Oxford, UK, OX1 3NP}
%
%\author{Ard A. Louis}
%\affiliation{Rudolf Peierls Centre for Theoretical Physics, Department of Physics, University of Oxford, 1 Keble Road, Oxford, UK, OX1 3NP}
%
%\author{Jonathan P. K. Doye}
%\affiliation{Physical \& Theoretical Chemistry Laboratory, Department of Chemistry, University of Oxford, South Parks Road, Oxford, UK, OX1 3QZ}
%
%\date{\today}
%
%\pacs{}

%\maketitle

\beginsupplement
\makeatletter
\renewcommand{\bibnumfmt}[1]{[S#1]}
\renewcommand{\citenumfont}[1]{S#1}

%\FloatBarrier
%\tableofcontents

%\FloatBarrier
\section{Simulation methods}
\subsection{Cyclization simulations}
\label{sec:cyclization-methods-cyclization-simulations}
Cyclization simulations were performed in three phases: exploratory, equilibration and production. In all cases, we use a virtual-move Monte Carlo (VMMC) algorithm \cite{Swhitelam_avoiding_2007} in combination with umbrella sampling \cite{Storrie_nonphysical_1977}.

In the exploratory phase, we iteratively adjusted the umbrella sampling bias for the windows associated with the open and cyclized states, yielding a flat population of states for both windows. We use discrete potentials for both dimensions of the order parameter (see main text). For $Q_\textnormal{ee}$, we use variable width distance increments: \SIrange{0}{1.7036}{\nano\meter}, 3.4072, 5.1108, 8.518, 12.777, 17.036, 21.295,  25.554, 34.072, 42.59, 51.108 and $> \SI{51.108}{\nano\meter}$. For $Q_\textnormal{bp}$, the increment is fixed, and is simply the number of base pairs formed ($0 \le Q_\textnormal{bp} \le N_\textnormal{s}$). As the weighting iterations were semi-automated, simulation times varied in the range \SIrange{E6}{E7} VMMC steps per particle. Simulations were performed with one molecule in a cubic box of dimension \SI{170.36}{\nano\meter}, which corresponds to a unimolecular concentration of \SI{336}{\nano\Molar}. To further simplify sampling, we forbid the formation of base pairs that are not intended in the design of the system (non-native base pairs).

In the equilibration phase, we equilibrated the system for \SI{E7}{} VMMC steps per particle using the aforementioned umbrella weights. Adequate sampling (number of transitions in $Q_\textnormal{ee}$ and $Q_\textnormal{bp}$) and decorrelation (via a block average decorrelation method) of potential energy, bubble size, fraying and structural kinking were checked. For the production phase, each simulation (five independent simulations per measurement) was initialized with a statistically independent starting configuration and a unique random seed. This is accomplished by randomly drawing starting configurations from the equilibration phase, one for each production trial, ignoring the first \SI{2E6}{} VMMC steps per particle of equilibration. The production trial run time is \SI{E7}{} VMMC steps per particle. For reference, the characteristic decorrelation time for the potential energy is $\sim \SI{E4}{}$ VMMC steps per particle, while the decorrelation time for kinking is $\sim \SI{E5}{}$ VMMC steps per particle.

The ``seed moves'' used to build clusters in the VMMC algorithm were:
\begin{itemize}
\item Rotation of a nucleotide about its backbone site, with an axis chosen uniformly on the unit sphere, and with an angle drawn from a normal distribution with a mean of zero and a standard deviation of 0.10 radians. 
\item Translation of a nucleotide, where the displacement along each Cartesian axis is drawn from a normal distribution with a mean of zero and a standard deviation of \SI{0.08518}{\nano\meter}. 
\end{itemize}

\afterpage{
\begin{table*}
\begin{tabularx}{\textwidth}{ llp{6in} }
\toprule
\multicolumn{1}{c}{$N_\textnormal{bp}$} & \multicolumn{1}{c}{Note} & \multicolumn{1}{c}{Cyclization sequence} \\
\midrule
\multicolumn{3}{l}{$N_\textnormal{s}=\SI{10}{\bp}$} \\
\midrule
$30^\ddagger$  &                                 & \tiny{\texttt{{\color{red}CAG AAT CCG T}GC TAC ACC TCC ACC GTT TCA}}\\
67  &                                 & \tiny{\texttt{{\color{red}CAG AAT CCG T}GC TAG TAC CTC AAT ATA GAC TCC CTT TGA CCT GAC TAT CCT CAC CTC CAC CGT TTC A}}\\
    &                                 & \tiny{\texttt{CGA TCA TGG AGT TAT ATC TGA GGG AAA CTG GAC TGA TAG GAG TGG AGG TGG CAA AGT {\color{red}GTC TTA GGC A}}}\\
69  &                                 & \tiny{\texttt{{\color{red}CAG AAT CCG T}GC TAG TAC CTC AAT ATA GAC TCC CTT TGA CCC ATG ACT ATC CTC ACC TCC ACC GTT TCA}} \\
71  &                                 & \tiny{\texttt{{\color{red}CAG AAT CCG T}GC TAG TAC CTC AAT ATA GAC TCC CTT CTA ATT GAC TGA CTA TCC TCA CCT CCA CCG TTT CA}} \\
73  &                                 & \tiny{\texttt{{\color{red}CAG AAT CCG T}GC TAG TAC CTC AAT ATA GAC TCC CTT CTA ATT GAC CAT GAC TAT CCT CAC CTC CAC CGT TTC A}} \\
74  &                                 & \tiny{\texttt{{\color{red}CAG AAT CCG T}GC TAG TAC CTC AAT ATA GAC TCC CTT TAA GTT GAC CCA TGA CTA TCC TCA CCT CCA CCG TTT CA}} \\
76  &                                 & \tiny{\texttt{{\color{red}CAG AAT CCG T}GC TAG TAC CTC AAT ATA GAC TCC CTT CTA AGT TGA CCT CAT GAC TAT CCT CAC CTC CAC CGT TTC A}} \\
93  &                                 & \tiny{\texttt{{\color{red}CAG AAT CCG T}GC TAG TAC CTC AAT ATA GAC TCC CTT AAT ACT TCT CCT ATG ACT TCT AAT TGA CCC ATG ACT ATC CTC ACC TCC ACC GTT TCA}} \\
94  &                                 & \tiny{\texttt{{\color{red}CAG AAT CCG T}GC TAG TAC CTC AAT ATA GAC TCC CTT TAA TAC TTC TCC TAT GAC TTC TAA TTG ACC CAT GAC TAT CCT CAC CTC CAC CGT TTC A}} \\
97  &                                 & \tiny{\texttt{{\color{red}CAG AAT CCG T}GC TAG TAC CTC AAT ATA GAC TCC CTA TGT TAA TAC TTC TCC TAT GAC TTC TAA TTG ACC CAT GAC TAT CCT CAC CTC CAC CGT TTC A}} \\
99  &                                 & \tiny{\texttt{{\color{red}CAG AAT CCG T}GC TAG TAC CTC AAT ATA GAC TCC CTA GAT GTT AAT ACT TCT CCT ATG ACT TCT AAT TGA CCC ATG ACT ATC CTC ACC TCC ACC GTT TCA}} \\
101  &                                 & \tiny{\texttt{{\color{red}CAG AAT CCG T}GC TAG TAC CTC AAT ATA GAC TCC CTG TAG ATG TTA ATA CTT CTC CTA TGA CTT CTA ATT GAC CCA TGA CTA TCC TCA CCT CCA CCG TTT CA}} \\
103  &                                 & \tiny{\texttt{{\color{red}CAG AAT CCG T}GC TAG TAC CTC AAT ATA GAC TCC CTA CGT AGA TGT TAA TAC TTC TCC TAT GAC TTC TAA TTG ACC CAT GAC TAT CCT CAC CTC CAC CGT TTC A}} \\
105  &                                 & \tiny{\texttt{{\color{red}CAG AAT CCG T}GC TAG TAC CTC AAT ATA GAC TCC CTG AAC GTA GAT GTT AAT ACT TCT CCT ATG ACT TCT AAT TGA CCC ATG ACT ATC CTC ACC TCC ACC GTT TCA}} \\
106  &                                 & \tiny{\texttt{{\color{red}CAG AAT CCG T}GC TAG TAC CTC AAT ATA GAC TCC CTA GAA CGT AGA TGT TAA TAC TTC TCC TAT GAC TTC TAA TTG ACC CAT GAC TAT CCT CAC CTC CAC CGT TTCA}} \\
207  &                                 & \tiny{\texttt{{\color{red}CAG AAT CCG T}GC TAG TAC CTC AAT ATA GAC TCC CTA TCA GTA CGA AGC TGG GCT ATA CCG TTC TTA TTG TCC TTA ATA CCA CCG ACG AGT TGT ACG CCC TCT CAT CCG AAG ACG ACA CGT ACC TGG GAA AAG AAC GTA GAT GTT AAT ACT TCT CCT ATG ACT TCT AAT TGA CCC ATG ACT ATC CTC ACC TCC ACC GTT TCA}} \\

\\ \multicolumn{3}{l}{$N_\textnormal{d}=\SI{91}{\bp}$} \\
\midrule
101  &                                 & \tiny{\texttt{{\color{red}CAG AAT CCG T}GC TAG TAC CTC AAT ATA GAC TCC CTG TAG ATG TTA ATA CTT CTC CTA TGA CTT CTA ATT GAC CCA TGA CTA TCC TCA CCT CCA CCG TTT CA}} \\
$100^\ddagger$  &                                 & \tiny{\texttt{{\color{red}AGA ATC CGT} GCT AGT ACC TCA ATA TAG ACT CCC TGT AGA TGT TAA TAC TTC TCC TAT GAC TTC TAA TTG ACC CAT GAC TAT CCT CAC CTC CAC CGT TTC A}} \\
$99^\ddagger$   &                                 & \tiny{\texttt{{\color{red}GAA TCC GT}G CTA GTA CCT CAA TAT AGA CTC CCT GTA GAT GTT AAT ACT TCT CCT ATG ACT TCT AAT TGA CCC ATG ACT ATC CTC ACC TCC ACC GTT TCA}} \\
$98^\ddagger$   &                                 & \tiny{\texttt{{\color{red}AAT CCG T}GC TAG TAC CTC AAT ATA GAC TCC CTG TAG ATG TTA ATA CTT CTC CTA TGA CTT CTA ATT GAC CCA TGA CTA TCC TCA CCT CCA CCG TTT CA}} \\
$97^\ddagger$   &                                 & \tiny{\texttt{{\color{red}ATC CGT} GCT AGT ACC TCA ATA TAG ACT CCC TGT AGA TGT TAA TAC TTC TCC TAT GAC TTC TAA TTG ACC CAT GAC TAT CCT CAC CTC CAC CGT TTC A}} \\
$96^\ddagger$   &                                 & \tiny{\texttt{{\color{red}TCC GT}G CTA GTA CCT CAA TAT AGA CTC CCT GTA GAT GTT AAT ACT TCT CCT ATG ACT TCT AAT TGA CCC ATG ACT ATC CTC ACC TCC ACC GTT TCA}} \\
$95^\ddagger$   &                                 & \tiny{\texttt{{\color{red}CCG T}GC TAG TAC CTC AAT ATA GAC TCC CTG TAG ATG TTA ATA CTT CTC CTA TGA CTT CTA ATT GAC CCA TGA CTA TCC TCA CCT CCA CCG TTT CA}} \\

\\ \multicolumn{3}{l}{Variable-sequence $N_\textnormal{d}$} \\
\midrule
73   & TA                              & \tiny{\texttt{{\color{red}CAG AAT CCG T}AG CTC TAG CAC CGC TTA AAC GCA CGT ACG CGC TGT CTA CCG CGT TTT AAC CGC CAA TAG GAT T}} \\
73   & E8A10                           & \tiny{\texttt{{\color{red}CAG AAT CCG T}TT TTA TTT ATC GCC TCC ACG GTG CTG TTT TTT TTT TCT GTT GGC CGT GTT ATC TCG AGT TAG T}} \\
73   & E8A17                            & \tiny{\texttt{{\color{red}CAG AAT CCG T}TT TTA TTT ATC GCC TCC ACG GTG CTT TTT TTT TTT TTT TTT GGC CGT GTT ATC TCG AGT TAG T}} \\
73   & E8A26                           & \tiny{\texttt{{\color{red}CAG AAT CCG T}TT TTA TTT ATC GCC TCC TTT TTT TTT TTT TTT TTT TTT TTT TTC CGT GTT ATC TCG AGT TAG T}} \\
73   & E8A38                            & \tiny{\texttt{{\color{red}CAG AAT CCG T}TT TTA TTT ATC GTT TTT TTT TTT TTT TTT TTT TTT TTT TTT TTT TTT TTT ATC TCG AGT TAG T}} \\

\\ \multicolumn{3}{l}{Structural defects} \\
\midrule
69$^\textnormal{m}$   & C:C mismatch              & \tiny{\texttt{{\color{red}CAG AAT CCG T}GC TAG TAC CTC AAT ATA GAC TCC CTT TGA {\color{blue}C}CC ATG ACT ATC CTC ACC TCC ACC GTT TCA}} \\
97$^\textnormal{n}$   & Nick 1                    & \tiny{\texttt{{\color{red}CAG AAT CCG T}GC TAG TAC CTC AAT ATA GAC TCC CTA {\color{blue}|} TGT TAA TAC TTC TCC TAT GAC TTC TAA TTG ACC CAT GAC TAT CCT CAC CTC CAC CGT TTC A}} \\
                      &                           & \tiny{\texttt{CGA TCA TGG AGT TAT ATC TGA GGG ATA CAA TTA TGA AGA GGA TAC TGA AGA TTA ACT GGG TAC TGA TAG GAG TGG AGG TGG CAA AGT {\color{red}GTC TTA GGC A}}} \\
97$^\textnormal{n}$   & Nick 2                    & \tiny{\texttt{{\color{red}CAG AAT CCG T}GC TAG TAC CTC AAT ATA GAC TCC CTA TGT TAA TAC TTC TCC TAT GAC TTC TAA TTG ACC CAT GAC TAT CCT CAC CTC CAC CGT TTC A}} \\
                      &                           & \tiny{\texttt{CGA TCA TGG AGT TAT ATC TGA GGG ATA CAA TTA TGA AGA GGA TAC TGA AGA TTA ACT GGG TAC {\color{blue}|} TGA TAG GAG TGG AGG TGG CAA AGT {\color{red}GTC TTA GGC A}}} \\
97$^\textnormal{n}$   & Double Nick               & \tiny{\texttt{{\color{red}CAG AAT CCG T}GC TAG TAC CTC AAT ATA GAC TCC CTA {\color{blue}|} TGT TAA TAC TTC TCC TAT GAC TTC TAA TTG ACC CAT GAC TAT CCT CAC CTC CAC CGT TTC A}} \\
                      &                           & \tiny{\texttt{CGA TCA TGG AGT TAT ATC TGA GGG ATA CAA TTA TGA AGA GGA TAC TGA AGA TTA ACT GGG TAC {\color{blue}|} TGA TAG GAG TGG AGG TGG CAA AGT {\color{red}GTC TTA GGC A}}} \\

\bottomrule
\end{tabularx}
\caption
{
Unless otherwise stated,$^\ddagger$ cyclization sequences are from V\&H \cite{Svafabakhsh_extreme_2012}, with sticky end ($N_\textnormal{s}$) highlighted in red, and the remainder of the sequence ($N_\textnormal{d}$) in black. 
Sequences are written $5^\prime$ to $3^\prime$. For conciseness, other than the initial example of $N_\textnormal{bp}=\SI{67}{\bp}$, a second strand is omitted where it is the reverse complement of the first strand. \newline
$^\ddagger$ Non-V\&H sequence. \newline
$^\textnormal{m}$ C:C Mismatch highlighted (\textcaption{blue}).\newline
$^\textnormal{n}$ Nicks highlighted with bar | (\textcaption{blue}).
}
\label{tab:cyclization-methods-sequences}
\end{table*}
}

\subsection{Dimerization simulations}
\label{sec:cyclization-methods-dimerization-simulations}
Dimerization simulations follow a very similar procedure as cyclization simulations (\Cref{sec:cyclization-methods-cyclization-simulations}), but adapted for a bimolecular system.  With the exception of the starting configuration (bimolecular), exploratory, equilibration and production procedures are identical. As discussed in the main text, each molecule has one complementary sticky end and one blunt end. This precludes the formation of multimers and circular dimers, allowing only the formation of linear dimers. Simulations were performed with these two complementary, non-palindromic, sticky-ended duplexes, in a cubic box of dimension \SI{170.36}{\nano\meter}, which corresponds to a dimer concentration of \SI{336}{\nano\Molar}. To estimate the impact of excluded volume effects, dimerization systems were also performed in a box of half the size and therefore $8$ times the concentration (\SI{85.18}{\nano\meter} per side, which corresponds to a dimer concentration of \SI{2.69}{\micro\Molar}). As for the cyclization simulations, non-native base pairing was disallowed. 

Analogously to the cyclization simulations, the dimerization simulations are windowed using the order parameters $Q_\textnormal{ee}$ and $Q_\textnormal{bp}$, respectively the distance of closest approach and the number of base pairs formed between complementary sticky ends. The window associated with the dimerized state is given by $Q_\textnormal{bp} \ge \SI{1}{bp}$ ($Q_\textnormal{ee} = Q^\textnormal{min}_\textnormal{ee}$, the small separation between base sites of the base paired nucleotides), while the window associated with the undimerized state is given by $Q_\textnormal{bp} = \SI{0}{bp}$.

\subsection{Computation of equilibrium constants}
\label{sec:cyclization-methods-equilibrium-constants}
The cyclization reaction is unimolecular (\Cref{fig:cyclization-diagram-system-oxDNA}\,(a)):
\begin{equation}
\ce{A} \ce{<=>[k_\textnormal{cyc}][k_\textnormal{uncyc}]} \ce{B},
\end{equation}
where \ce{A} and \ce{B} are the open and cyclized states respectively, $k_\textnormal{cyc}$ is the forward rate constant (cyclization) and $k_\textnormal{uncyc}$ is the reverse rate constant (uncyclization). The equilibrium constant for cyclization, $K_\textnormal{eq}^\textnormal{cyc}$, can be estimated directly from simulations of a unimolecular, isolated system as 
\begin{equation}
K_\textnormal{eq}^\textnormal{cyc} = \frac{ P_\ce{B} }{ P_\ce{A} }.
\end{equation}
Here $P_\ce{A}$ and $P_\ce{B}$ are the probabilities with which uncyclized and cyclized systems are observed in simulation ($P_\ce{A} + P_\ce{B} = 1$).

The dimerization reaction is a bimolecular association of distinct molecules:
\begin{equation}
\ce{A} + \ce{B} \ce{<=>[k_\textnormal{dim}][k_\textnormal{undim}]} \ce{AB}.
\end{equation}
In this case, the bimolecular equilibrium constant can be inferred from a simulation of a single pair of dimerizing monomers via
\begin{align}
K_\textnormal{eq}^\textnormal{dim} &= \frac{ P_\ce{AB} }{ P_{\rm A/B} \ce{[A_0]} },
\end{align}
in which $[A_0]$ is the total concentration of strand type A in simulation, and $P_{\rm A/B}$ and $P_\ce{AB}$ are the probabilities with which monomers and dimers are observed in simulation. This expression follows from Eq. 7 of Ref.\ \cite{ouldridge_extracting_2010}. The resultant $K_\textnormal{eq}^\textnormal{dim}$ obeys the standard relation for bulk systems in equilibrium,
\begin{align}
K_\textnormal{eq}^\textnormal{dim} &= \frac{ [\ce{AB}] }{[\ce{A}] [\ce{B}] }.
\end{align}

Recall from the main text that we use a definition for the equilibrium $j$-factor $j_\textnormal{eq}$ that is consistent with Vafabakhsh and Ha (V\&H) \cite{Svafabakhsh_extreme_2012}:
\begin{equation}
j_\textnormal{eq} \equiv \frac{ K_\textnormal{eq}^\textnormal{cyc} }{ K_\textnormal{eq}^\textnormal{dim} }. 
\end{equation}
This is a measure of the effective concentration of one sticky end in the vicinity of the other when the dimerization reaction involves distinct monomers, forming a heterodimer, with only one sticky end per monomer. Researchers often estimate $K_\textnormal{eq}^\textnormal{dim}$ using the dimerization of identical monomers, forming a homodimer, with palindromic sticky ends. In this case, given the same underlying interaction strength between sticky ends (and hence the same $K_\textnormal{eq}^\textnormal{cyc}$), $K_\textnormal{eq}^\textnormal{dim} $ would be increased by a factor of two due to combinatorial (4 times the number of possible dimers) and symmetry effects (a factor of 1/2 for a homodimer rather than a heterodimer) \cite{taylor_application_1990, ouldridge_extracting_2010}. As a result, with this approach the equilibrium $j$-factor is estimated as 
\begin{equation}
j_\textnormal{eq} = 2 \frac{ K_\textnormal{eq}^\textnormal{cyc} }{ K_\textnormal{eq}^\textnormal{dim} }. 
\end{equation}

In the case of identical monomers with two non-palindromic sticky ends, there is no pre-factor in the $j$-factor estimate, as the combinatorial (twice the number of possible dimers) and symmetry effects (a factor of 1/2 for a homodimer) cancel. Overall, under the approximation of ideal behaviour of separate complexes, these three approaches are equivalent.

\subsection{Structural criterion for kink detection}
\label{sec:cyclization-methods-structural-kink-criterion}
While it is often visually straightforward to identify kinks, automating their detection is not without difficulty. We have developed both energetic and structural criteria for identifying kinks, which are based on disruption of stacking interactions and changes in base orientation, respectively. In this study, we rely on the structural criterion; we have discussed the differences between the two criteria in detail in the supplementary material of the accompanying paper \cite{SHarrison_Cohen}. 

Structurally, we define a kink using the relative orientation of consecutive nucleotides. In oxDNA, the orientation of each nucleotide is unequivocally determined by two orthogonal unit vectors, the base-backbone vector and the base-normal vector. The base-backbone vector connects the backbone and base interaction sites of each nucleotide. The model is designed such that, in a relaxed duplex, the base-normal vector at index $i$, $\mathbf{\hat{a}}_i$, and at the consecutive index $i+1$, $\mathbf{\hat{a}}_{i+1}$, are approximately parallel; that is, $\mathbf{\hat{a}}_i \cdot \mathbf{\hat{a}}_{i+1} \approx 1$. A kink is defined to be present if $\mathbf{\hat{a}}_i \cdot \mathbf{\hat{a}}_{i+1} < 0$, a condition implying a more than 90$^\circ$ change in orientation of consecutive nucleotides along one strand. 
Naturally, for duplex regions, we consider kinking along either strand; although, usually if a kink is present in the duplex, then both strands are kinked. To limit false positives due to fraying, we do not include the first and last 3 pairs of nucleotides in the duplex region in our analysis.

In the fully cyclized configuration there are two ``nicks'' where the two ends of each strand meet. 
Kinks in nicked regions are treated slightly differently, as our criterion is only well defined along an intact strand. A kink at a nick is detected on the opposite strand. As kinks at a nick may diffuse slightly, we define a region of \SI{3}{base pairs} on either side of the nick on the intact strand. If the intact strand is kinked in this \SI{6}{base pair} region, then the molecule is considered kinked at that nick.

Since distinguishing fraying from kinking at low $Q_\textnormal{bp}$ is problematic, we compute kinking only for the most probable values, e.g.\ $Q_\textnormal{bp}=\SIrange{8}{10}{\bp}$ for the $N_\textnormal{s}=\SI{10}{\bp}$ molecules depicted in \Cref{fig:cyclization-kink}. The lower $Q_\textnormal{bp}$ configurations appear in our simulations only because of biased sampling and have a negligible contribution to the equilibrium kinking probability.

\section{Additional results}
\subsection{Dimerization equilibrium}
\label{sec:cyclization-Keq-dimerization}
The equilibrium constant for dimerization $K_\textnormal{eq}^\textnormal{dim}$ is length-independent to within two times the standard error of the mean for the oxDNA average-base parameterization (\Cref{tab:cyclization-Keq-dimerization}). Varying concentration suggests that the role of excluded volume effects is small between \SI{336}{\nano\Molar} and \SI{2.69}{\micro\Molar}.

For the sequence-dependent parameterization, there is some variation in $K_\textnormal{eq}^\textnormal{dim}$, but this does not represent length-dependence per-se, rather it is most likely due to sequence variation in the bases at the interface between the duplexes and the sticky ends.

For the computation of $j_\textnormal{eq}$ and $j_\textnormal{dyn}$, we use average values for \SI{336}{\nano\Molar}. For sequence-dependent results at $N_\textnormal{bp}=\SI{73}{\bp}$, a length where the dimerization equilibrium constant was computed, we use the appropriate length-specific value instead of the average.

\begin{table*}
\ra{1.3}
\begin{tabularx}{\textwidth}{p{0.5in}p{0.5in} ccc p{0.1in} ccc}
\toprule
\multicolumn{1}{l}{$N_\textnormal{d} / \SI{}{bp}$} & 
\multicolumn{1}{l}{$\ce{[Conc]} / \SI{}{\nano\Molar}$} & 
\multicolumn{3}{c}{Average-base} & 
\multicolumn{1}{l}{} &
\multicolumn{3}{c}{Sequence-dependent} \\
\cmidrule(l){3-5} \cmidrule(r){7-9}
\multicolumn{1}{l}{} & 
\multicolumn{1}{l}{} & 
\multicolumn{1}{l}{$K_\textnormal{eq}^\textnormal{dim}$ / \SI{e12}{\per\Molar}} &
\multicolumn{1}{l}{$\Delta G_\textnormal{dim}^\ddagger$ / \SI{}{\kT}} &
\multicolumn{1}{l}{$\Delta G_\textnormal{undim}^\ddagger$ / \SI{}{\kT}} &
\multicolumn{1}{c}{} & 
\multicolumn{1}{l}{$K_\textnormal{eq}^\textnormal{dim}$ / \SI{e12}{\per\Molar}} & 
\multicolumn{1}{l}{$\Delta G_\textnormal{dim}^\ddagger$ / \SI{}{\kT}} & 
\multicolumn{1}{l}{$\Delta G_\textnormal{undim}^\ddagger$ / \SI{}{\kT}} \\
\midrule
20;20  & 336   & $1.11 \pm 0.11$ & $14.42 \pm 0.11$ & $27.09 \pm 0.10$  & & $0.42 \pm 0.05$ & $14.13 \pm 0.07$ & $25.75 \pm 0.11$ \\
57;57  & 336   & $0.87 \pm 0.05$ & $14.62 \pm 0.04$ & $27.05 \pm 0.05$  & & $0.54 \pm 0.03$ & $14.01 \pm 0.01$ & $25.79 \pm 0.06$ \\
63;63  & 336   & $0.78 \pm 0.04$ & $14.69 \pm 0.15$ & $27.00 \pm 0.05$  & & $0.33 \pm 0.05$ & $14.19 \pm 0.05$ & $25.60 \pm 0.13$ \\
91;91  & 336   & $0.91 \pm 0.15$ & $14.66 \pm 0.07$ & $27.13 \pm 0.16$  & & $0.44 \pm 0.03$ & $14.20 \pm 0.04$ & $25.86 \pm 0.08$ \\
\midrule
Avg    & 336   & $0.92 \pm 0.06$ & $14.60 \pm 0.06$ & $27.07 \pm 0.05$  & & $0.43 \pm 0.03$ & $14.12 \pm 0.03$ & $25.75 \pm 0.05$ \\ \\

10;10  & 2690  & $1.22 \pm 0.19$ & $12.54 \pm 0.05$ & $27.38 \pm 0.16$  & & $0.46 \pm 0.02$ & $12.11 \pm 0.11$ & $25.91 \pm 0.04$ \\
28;29  & 2690  & $1.11 \pm 0.09$ & $12.66 \pm 0.10$ & $27.41 \pm 0.08$  & & $0.82 \pm 0.12$ & $12.60 \pm 0.04$ & $27.03 \pm 0.15$ \\
30;33  & 2690  & $1.11 \pm 0.17$ & $12.64 \pm 0.12$ & $27.40 \pm 0.15$  & & $1.15 \pm 0.12$ & $12.53 \pm 0.05$ & $27.32 \pm 0.11$ \\
45;46  & 2690  & $1.04 \pm 0.09$ & $12.55 \pm 0.08$ & $27.24 \pm 0.09$  & & $0.48 \pm 0.05$ & $12.08 \pm 0.04$ & $25.90 \pm 0.11$ \\
\midrule
Avg    & 2690  & $1.12 \pm 0.07$ & $12.60 \pm 0.05$ & $27.36 \pm 0.06$  & & $0.73 \pm 0.08$ & $12.32 \pm 0.06$ & $26.54 \pm 0.15$ \\
\bottomrule
\end{tabularx}
\caption[]
{
Equilibrium constant for dimerization ($K_\textnormal{eq}^\textnormal{dim}$)
and activation free-energy barriers to dimerization ($\Delta
G_\textnormal{dim}^\ddagger$) and undimerization ($\Delta
G_\textnormal{undim}^\ddagger$) as defined in the main text
for both the oxDNA average-base
and sequence-dependent parameterizations. Simulations were performed at
$T=\SI{298}{\kelvin}$. 
The first column gives the lengths of duplex regions of the two monomers. The complementary sticky ends are of identical sequence and 
length $N_\textnormal{s}=\SI{10}{\bp}$. 
Sequences from \Cref{tab:cyclization-methods-sequences} (for
\ce{[\textnormal{Conc}]=\SI{336}{\nano\Molar}} the relevant sequences are those with $N_\textnormal{bp}=N_\textnormal{d}+10$;
for \ce{[\textnormal{Conc}]=\SI{2690}{\nano\Molar}} they are those with $N_\textnormal{bp}$ equal to the total length of the dimer).
\newline
}
\label{tab:cyclization-Keq-dimerization}
\end{table*}

\subsection{Initial stress in cyclized system}
\label{sec:cyclization-diagram-Qbp1}
\begin{figure}
\includegraphics[width=8.5cm]{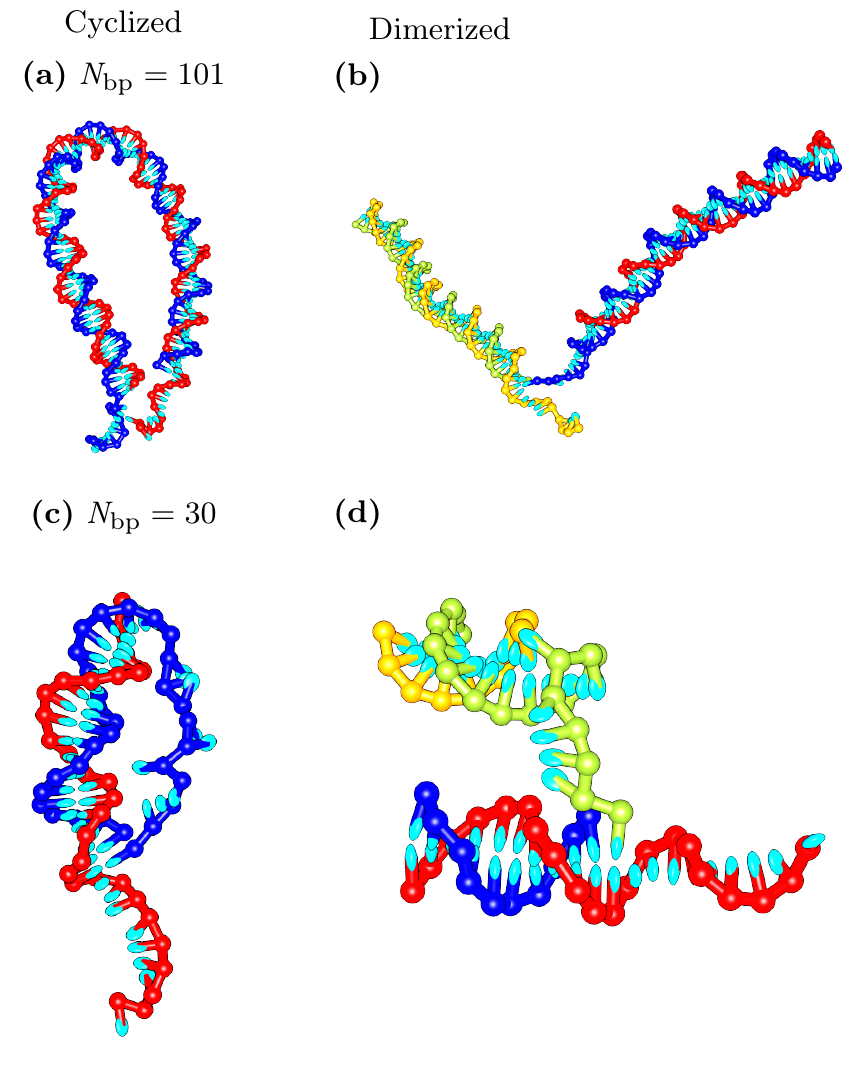}
\caption[]
{
OxDNA representations of $Q_\textnormal{bp}=\SI{1}{\bp}$ states for cyclized and dimerized configurations, highlighting the bending required to form the first base pair. Dimerized configurations in (b) and (d) are shown for $N_\textnormal{d1} + N_\textnormal{d2} + N_\textnormal{s}=N_\textnormal{bp}$, instead of length $N_\textnormal{bp}$ monomers. For (b), $N_\textnormal{d1}=45$, $N_\textnormal{d2}=46$ and $N_\textnormal{s}=10$. For (d), $N_\textnormal{d1}=N_\textnormal{d2}=10$ and $N_\textnormal{s}=10$.
}
\label{fig:cyclization-diagram-Qbp1}
\end{figure}

There is a general trend towards a larger activation free-energy barrier to cyclization ($\Delta G_\textnormal{cyc}^\ddagger$) at shorter $N_\textnormal{bp}$. This is due to the bending stress imposed upon the system by the formation of the initial base pair ($Q_\textnormal{bp}=1$). At the very shortest lengths $N_\textnormal{bp} \approx \SIrange{30}{45}{\bp}$, this trend is reversed because the complementary single-stranded sticky ends (of length $N_\textnormal{s}=\SI{10}{\bp}$) are sufficiently long relative to $N_\textnormal{d}$ that an initial base pair can form with less bending stress in the duplex region. States at the top of the free-energy barrier (i.e.\ $Q_\textnormal{bp}=1$) for cyclization and dimerization at $N_\textnormal{bp}=30,101$ are compared in \Cref{fig:cyclization-diagram-Qbp1}, clearly showing bending in the cyclized molecules, but not the dimerized molecules. Importantly, at $N_\textnormal{bp}=\SI{30}{\bp}$, the relatively long single-stranded region ($N_\textnormal{s}=\sfrac{1}{2} \times N_\textnormal{d} = 10$) reduces the requirement to bend the duplex (\Cref{fig:cyclization-diagram-Qbp1}(c)), compared to the strong bending necessary at $N_\textnormal{bp}=\SI{101}{\bp}$ (\Cref{fig:cyclization-diagram-Qbp1}(a)).

\subsection{Comparison to FRET experiments}
\label{sec:cyclization-experimental-Ha}
\subsubsection{Length of complementary sticky ends}
\label{sec:cyclization-experimental-Ha-Ns}
We compare oxDNA to available V\&H equilibrium data \cite{Svafabakhsh_extreme_2012} for $N_\textnormal{d}=\SI{91}{\bp}$, $N_\textnormal{s}=\SIrange{8}{10}{\bp}$. Specifically, V\&H report the fractional occupancy of the cyclized state $f_\textnormal{cyc}$. Akin to the nomenclature in \Cref{sec:cyclization-methods-equilibrium-constants}, $f_\textnormal{cyc} + f_\textnormal{open} = 1$, where $f_\textnormal{open}$ is the fractional occupancy of the open (uncyclized) state.

\begin{figure}
\includegraphics[width=8.5cm]{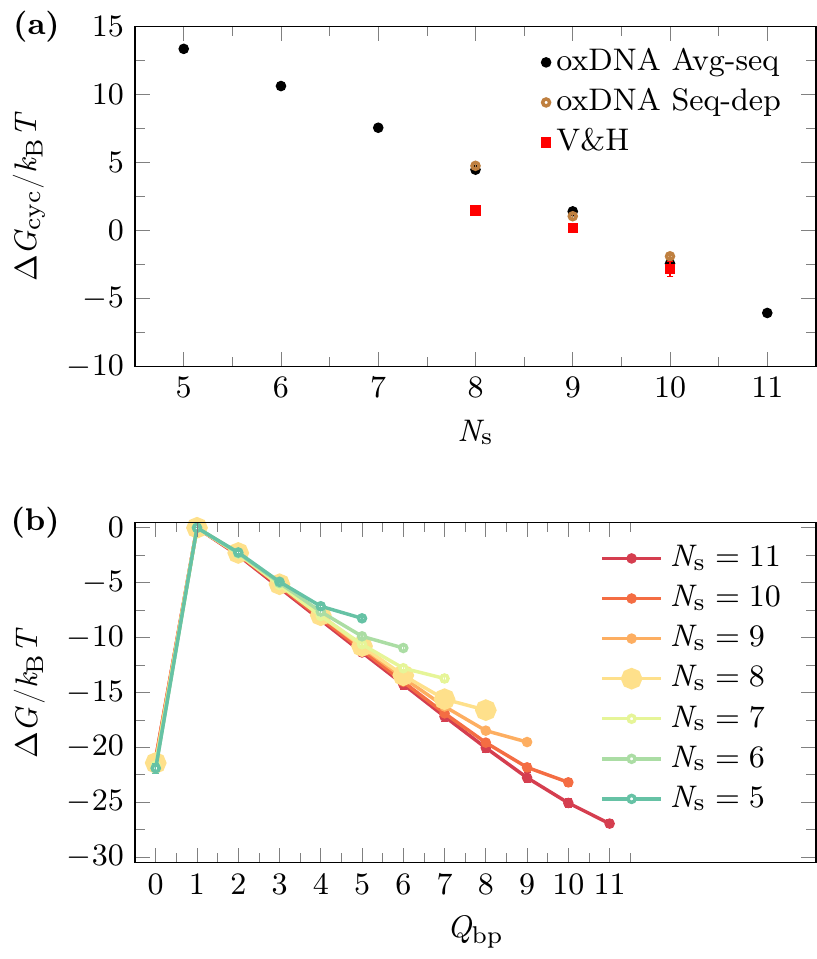}
\caption[]
{Dependence of cyclization thermodynamics on $N_\textnormal{s}$.
\textbf{(a)} Comparison between oxDNA and V\&H \cite{Svafabakhsh_extreme_2012} for $\Delta G_\textnormal{cyc}$, the free energy of the cyclized compared to the open state.
\textbf{(b)} Free-energy profiles for cyclization. 
The reference state corresponding to $\Delta G = 0$ is set to be 
at $Q_\textnormal{bp}=\SI{1}{\bp}$. 
All results are for $N_\textnormal{d}=91$, and using the oxDNA average-base parameterization, unless stated.
}
\label{fig:cyclization-experimental-Ha-Ns}
\end{figure}

Fractional occupancy can be a somewhat misleading measure by which to compare systems, because the entire transition region $f_\textnormal{cyc} \approx \SIrange{0.25}{0.75}{}$ is covered by a free-energy difference of $\sim \SI{2}{\kT}$, while a similar free-energy difference at the extrema ($f_\textnormal{cyc} \to 0 \textnormal{ or } 1$) would be indistinguishable. We therefore make our equilibrium comparison in terms of $\Delta G_\textnormal{cyc}$, where
\begin{equation}
\Delta G_\textnormal{cyc} = - k_\textnormal{B} T \ln \left( K_\textnormal{eq}^\textnormal{cyc} \right),
\end{equation}
and 
\begin{equation}
K_\textnormal{eq}^\textnormal{cyc} = \frac{ f_\textnormal{cyc} }{ f_\textnormal{open} }
                                   = \frac{ f_\textnormal{cyc} }{ 1- f_\textnormal{cyc} }.
\label{eq:cyclization-Keq-fcyc}
\end{equation}

OxDNA appears to be in good agreement with experiment; although, this may be partly coincidental as V\&H do not report either the temperature (presumed to be \SI{298}{\kelvin}) or a salt concentration (V\&H give their imaging buffer cation concentration as $\ce{[Na+]} = \SIrange{500}{1000}{\milli\Molar}$ or $\ce{[Mg^2+]} = \SIrange{10}{30}{\milli\Molar}$; recall that oxDNA is parameterized to $\ce{[Na+]} = \SI{500}{\milli\Molar}$). At $N_\textnormal{s}=\SI{9}{\bp}$ and $N_\textnormal{s}=\SI{10}{\bp}$, oxDNA and experiment differ by $\sim \SI{1}{\kT}$, while at $N_\textnormal{s}=\SI{8}{\bp}$, oxDNA under-reports experiment by $\sim \SI{3}{\kT}$ (cyclized state is less likely in oxDNA than experiment).

As expected from the SantaLucia model \cite{Ssantalucia_unified_1998, Ssantalucia_thermodynamics_2004} of nearest-neighbour thermodynamics, the linear relationship between oxDNA results in \Cref{fig:cyclization-experimental-Ha-Ns}\,(a) is reflective of the $\sim \SI{3}{\kT}$ stabilization of each additional base pair. 
A possible explanation for the discrepancy between oxDNA and V\&H \cite{Svafabakhsh_extreme_2012} is that hairpin formation in the $N_\textnormal{s}$ region could decrease the fraction of cyclized molecules; but given the sequence, hairpin formation should be negligible. In particular, we cannot explain the discrepancy between oxDNA and the V\&H data for $N_\textnormal{s}=\SI{8}{\bp}$, although we note that identifying a very low yield in experiment can be challenging, for instance due to the presence of impurities inducing large changes in $\Delta G_\textnormal{cyc}$.

\subsubsection{Nicks and mismatches}
\label{sec:cyclization-experimental-Ha-nick-mismatch}
\begin{figure}
\includegraphics[width=8.5cm]{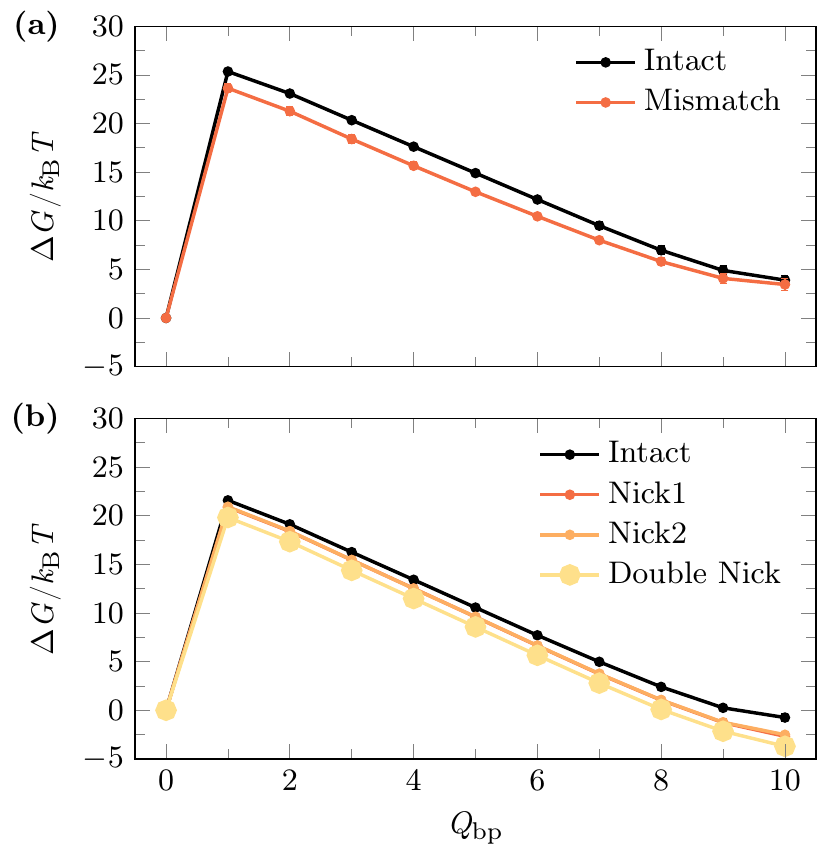}
\caption[]
{
Free-energy profiles for cyclization for \textbf{(a)} an intact duplex compared to a duplex with a single mismatch at $N_\textnormal{bp}=\SI{69}{\bp}$, and \textbf{(b)} intact versus nicked duplexes at $N_\textnormal{bp}=\SI{97}{\bp}$, using the oxDNA average-base parameterization and $N_\textnormal{s}=\SI{10}{\bp}$ for all sequences (\Cref{tab:cyclization-methods-sequences}).
}
\label{fig:cyclization-experimental-Ha-nick-mismatch}
\end{figure}

We examine the role of structural defects, namely nicks and mismatches, on cyclization using the average-base oxDNA parameterization. Unsurprisingly, we find that nicked and mismatches molecules cyclized more readily than their intact counterparts (\Cref{fig:cyclization-experimental-Ha-nick-mismatch}).

V\&H briefly examine the impact of nicks and mismatches on the rate of cyclization, recording $f_\textnormal{cyc}$ versus time (Figure S3 in Ref.\ \cite{Svafabakhsh_extreme_2012}). It does not appear that all molecules have reached an equilibrium plateau in $f_\textnormal{cyc}$, therefore a concrete equilibrium comparison to oxDNA is difficult. We do, however, observe good agreement between the oxDNA and the experimental observations of V\&H regarding the stabilizing effect of various motifs (\Cref{tab:cyclization-experimental-Ha-nick-mismatch}). However, we do not reproduce the unexpected similarity between the experimental values for intact sequences at $N_\textnormal{bp}=\SI{69}{\bp}$ and $N_\textnormal{bp}=\SI{97}{\bp}$. This is consistent with our inability to reproduce V\&H's $j$-factors for their shortest sequences.

\begin{table}
\begin{tabular}{ l c c c c c }
\toprule
\multicolumn{1}{l}{} &
\multicolumn{2}{c}{OxDNA} &
\multicolumn{1}{l}{} &
\multicolumn{2}{c}{Experiment$^\textnormal{*}$} \\
\cmidrule(l){2-3} \cmidrule(l){5-6}
\multicolumn{1}{l}{} &
\multicolumn{1}{c}{$K_\textnormal{eq}^\textnormal{cyc}$} &
\multicolumn{1}{c}{$\Delta\Delta G/\SI{}{\kT}$$^\dagger$} &
\multicolumn{1}{l}{} &
\multicolumn{1}{c}{$K_\textnormal{eq}^\textnormal{cyc}$} &
\multicolumn{1}{c}{$\Delta\Delta G/\SI{}{\kT}$$^\dagger$} \\
\midrule
$N_\textnormal{bp}=\SI{69}{\bp}$ \\
Intact      & $0.029 \pm 0.007$ &      && 1.10 \\
Mismatch    & $0.052 \pm 0.016$ & -0.6 && 3.14 & -1.1 \\
\\
$N_\textnormal{bp}=\SI{97}{\bp}$ \\
Intact      & $2.99 \pm 0.35$ &      && 0.93 \\
Nick 1      & $18.5 \pm 0.9$ & -1.8 && 1.73 & -0.6 \\
Nick 2      & $16.2 \pm 1.0$ & -1.7 \\
Double Nick & $50.3 \pm 2.4$ & -2.8 && 8.78 & -2.3 \\
\bottomrule
\end{tabular}
\caption[]
{
$K_\textnormal{eq}^\textnormal{cyc}$ for oxDNA and experiment (V\&H \cite{Svafabakhsh_extreme_2012}). The experimental assays were not intended to probe the equilibrium behaviour of the systems; as such, the last point of kinetic cyclization versus time assay may not represent equilibrium values. To emphasize the qualitative difference between intact molecules and those with structural defects (nicks, mismatches), oxDNA results are for the average-base parameterization. 
\newline
$^\textnormal{*}$ Converted from yield $f_\textnormal{cyc}$ to $K_\textnormal{eq}^\textnormal{cyc}$ using \Cref{eq:cyclization-Keq-fcyc}. 
\newline
$^\dagger$ $\Delta \Delta G$ is computed as the difference in the $\Delta G$ of cyclization between intact states and those with structural defects (nick or mismatch).
}
\label{tab:cyclization-experimental-Ha-nick-mismatch}
\end{table}

\subsection{Comparison to ligase experiments}
\label{sec:cyclization-experimental-Du-CW}
\begin{figure}
\includegraphics[width=8.5cm]{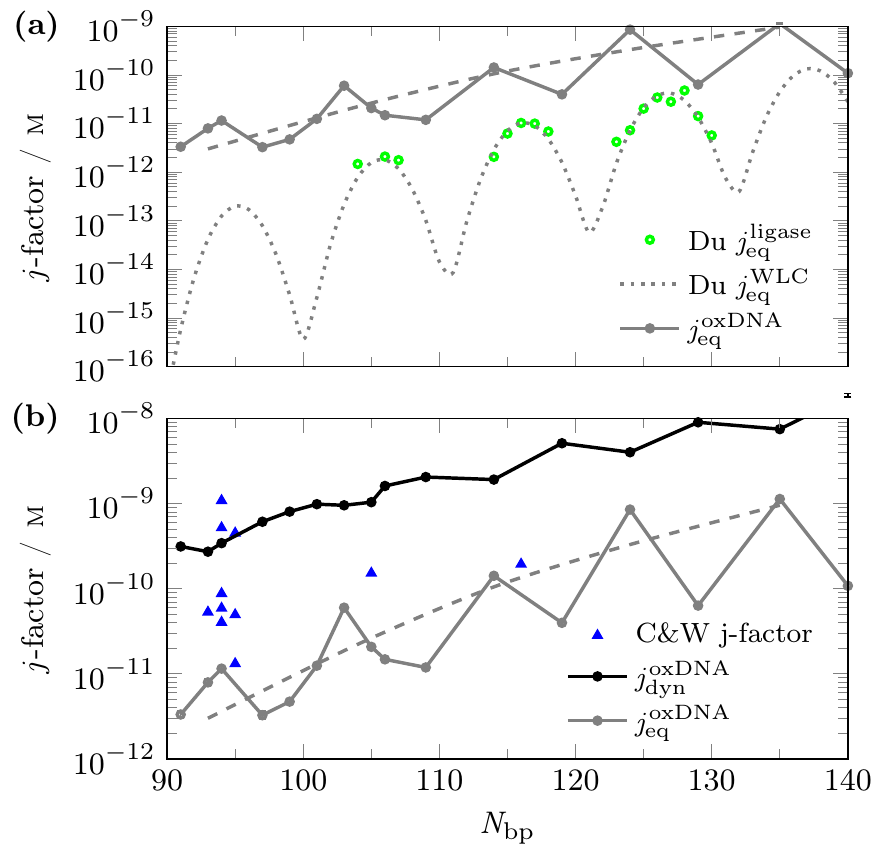}
\caption[]
{
OxDNA comparison to the experiments of Du \etal \cite{Sdu_cyclization_2005} and Cloutier \& Widom \cite{Scloutier_spontaneous_2004}.
\textbf{(a)} Du \etal equilibrium $j$-factor via a ligase-based assay $j_\textnormal{eq}^\textnormal{ligase}$ (\textcaption{green circles}) compared to the SY WLC expression $j_\textnormal{eq}^\textnormal{WLC}$ with Du \etal parameters (\textcaption{gray dotted}) and the oxDNA model $j_\textnormal{eq}^\textnormal{oxDNA}$ (\textcaption{gray marks}). The top envelop of $j_\textnormal{eq}^\textnormal{WLC}$ with oxDNA parameterized is dashed (\textcaption{gray dashed})
\textbf{(b)} C\&W ligase-based $j$-factor compared to $j_\textnormal{dyn}^\textnormal{oxDNA}$ (\textcaption{black marks}) and $j_\textnormal{eq}^\textnormal{oxDNA}$ (\textcaption{gray marks}). The maxima envelope of $j_\textnormal{eq}^\textnormal{oxDNA}$ is given with a gray dashed line.
}
\label{fig:cyclization-experimental-Du-CW}
\end{figure}

Comparison to the ligase experiments of Du \etal \cite{Sdu_cyclization_2005} and
Cloutier \& Widom (C\&W) \cite{Scloutier_spontaneous_2004} highlight some
interesting features of the oxDNA model
(\Cref{fig:cyclization-experimental-Du-CW}). For example, while both the Du
\etal experimental and oxDNA model results are similar to the SY WLC expression
for $N_\textnormal{bp} \gtrsim \SI{100}{\bp}$, there is a significant absolute
deviation in $j$-factor due to differences in the relevant mechanical
properties for oxDNA and those seemingly appropriate for the conditions of the
experiment of Du \etalnoperiod. Specifically, they use NEB T4-ligase buffer,
which has a salt concentration of $\ce{[Mg^2+]} = \SI{10}{\milli\Molar}$, while
oxDNA is parametrized to $\ce{[Na+]} = \SI{500}{\milli\Molar}$. Du
\etalnospace's fit to their data gives a weaker torsional stiffness
(\SI{2.4e-28}{} versus \SI{4.75e-28}{\joule\meter}), longer persistence
length (\SI{47}{} versus \SI{41.82}{\nano\meter}) and longer pitch length
(\SI{10.54}{} versus \SI{10.36}{bp/turn}) than for oxDNA. The comparison also
highlights how relatively small deviations in DNA mechanical properties (in particular a $\sim
\SI{10}{\%}$ change in the persistence length) may shift the apparent
$j_\textnormal{eq}$ by an order of magnitude.
The differences in SY WLC expression parameters are reasonable given the different solution conditions (the persistence length decreases
with increasing salt concentration \cite{Ssavelyev_monovalent_2012,herrerogalan13} and temperatures (e.g.\ $j_\textnormal{eq}$ is enhanced by a factor of $\sim 10$ between \SI{5}{\celsius} and \SI{42}{\celsius} \cite{geggier_temperature_2011}).
We note that the oxDNA persistence length is reasonable for the high salt conditions used by V\&H \cite{herrerogalan13}.

Du \etal only investigated lengths near the maxima in
$j_\textnormal{eq}^\textnormal{WLC}$. As a consequence, their experiments are
unable to confirm our prediction of a decrease in the magnitude of the
oscillations in $j_\textnormal{eq}^\textnormal{oxDNA}$, which we attribute to
the affect of teardrop configurations. Whether the experiments could
potentially see the effects of teardrop configurations also depends on how the
ligase activity depends on the nature of the DNA confirmation. Indeed,
Vologodskii \etal mention the possibility of poor ligation efficiency for
teardrop configurations \cite{Svologodskii_strong_2013}. 
We also note that because of the potential role of teardrop configurations, the
torsional modulus extracted from fits to the SY expression should be
interpreted with caution, and it is interesting to note that the values
typically obtained from these fits
\cite{Sshore_energetics_1983,taylor_application_1990,Sdu_cyclization_2005} are
significantly lower than those from single-molecule experiments with magnetic
tweezers \cite{Bryant2003,janssentorque}.

Since it is likely that C\&W conducted their experiment at too high a ligase concentration, it is difficult to characterize their $j$-factor in terms of either $j_\textnormal{dyn}$ or $j_\textnormal{eq}$. To isolate high ligase concentration as the culprit behind an apparent deviation from $j_\textnormal{eq}^\textnormal{WLC}$, Du \etal used conditions identical to C\&W, namely buffer ($\ce{[Mg^2+]} = \SI{10}{\milli\Molar}$) and temperature ($T=\SI{21}{\celsius}$ for Du \etalnospace, $T=\SI{20}{\celsius}$ for C\&W and $T=\SI{25}{\celsius}$ for oxDNA). Intriguingly, C\&W's results mainly lie inbetween $j_\textnormal{dyn}^\textnormal{oxDNA}$ and $j_\textnormal{eq}^\textnormal{oxDNA}$.

\subsection{Capture radius}
\label{sec:cyclization-capture-radius}
Comparing results for $j_\textnormal{dyn}$ with a WLC model requires that the finite dimensions of the sticky ends be taken into account. Defining a ``capture radius'', a measure of the distance within which the sticky ends can hybridize, is a common way of accounting for the finite size of DNA sticky ends. As oxDNA naturally accounts for the finite-size of the sticky ends, we have the ability to test the capture radius assumption.

In oxDNA, we define the capture radius $R_\textnormal{capture}$ as the distance between the first and last bases of the duplex region ($N_\textnormal{d}$) at $Q_\textnormal{bp}=\SI{1}{\bp}$ (\Cref{fig:cyclization-diagram-Qbp1}\,(a,c)). Nucleotide positions are used to report distance, instead of the mid-point along the helical axis, because fraying may occur at the ends of the duplex. The choice of nucleotide strand does not impact our results. 

\begin{figure}
\includegraphics[width=8.5cm]{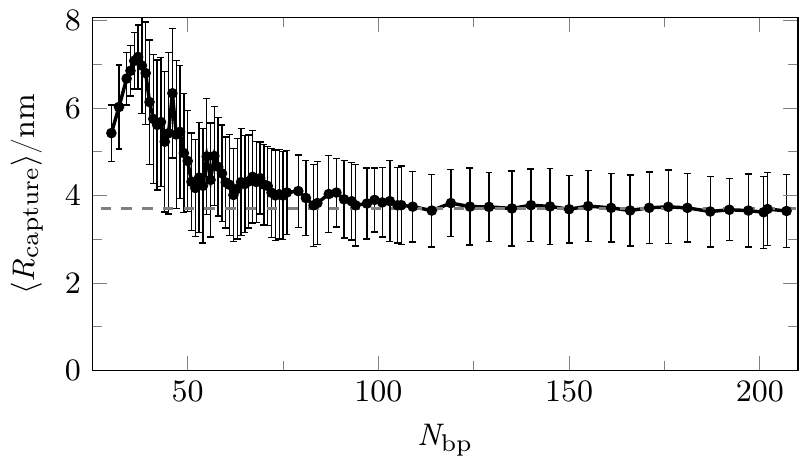}
\caption[]
{
The ensemble average of the oxDNA capture radius $\langle R_\textnormal{capture} \rangle$ as a function of length $N_\textnormal{bp}$ at fixed $N_\textnormal{s}=\SI{10}{\bp}$. 
Error bars represent the standard deviation.
}
\label{fig:cyclization-capture-radius}
\end{figure}

Our ensemble average capture radius, $\langle R_\textnormal{capture} \rangle$
is shown in \Cref{fig:cyclization-capture-radius}.  $\langle
R_\textnormal{capture} \rangle$ is roughly constant at $N_\textnormal{bp}
\approx \SIrange{100}{200}{\bp}$, increases slowly between $N_\textnormal{bp}
\approx \SIrange{50}{100}{\bp}$, and then more rapidly for $N_\textnormal{bp}
\approx \SIrange{40}{50}{\bp}$, before decreasing for $N_\textnormal{bp}
\lesssim \SI{37}{\bp}$.  The saturation value ($\langle R_\textnormal{capture}
\rangle\approx \SI{4}{\nano\meter}$) is quite close to the capture radius of
\SI{5}{\nano\meter} that has been used to interpret V\&H's experiments with
10-base sticky ends \cite{Svafabakhsh_extreme_2012,Svologodskii_bending_2013}. 

The free-energy cost associated with forming the initial contact is partitioned between that for bending the duplex and that for stretching the single-stranded sticky-ends.  At shorter $N_\textnormal{bp}$, the amount of bending required for initial binding increases, as does the stretching of the single-stranded sticky-ends. At short lengths, this trend is reversed when the length of the unperturbed duplex approaches the capture radius at $Q_\textnormal{bp}=\SI{1}{\bp}$. Interestingly, the distribution of $R_\textnormal{capture}$ is roughly Gaussian, with similar standard deviation for all lengths.

While we explicitly disallow misbonding, allowing misbonding may slightly
increase the capture radius, an effect that should be more prominent for longer
sticky-ends.  Overall, the \SI{5}{\nano\meter} value for the capture radius
used in the interpretation of V\&H's experiments appears not unreasonable.

\end{document}